\pdfoutput=1
\newif\ifconfver
\confvertrue        %declaring conference version true

\newif\ifonecoltab
\onecoltabtrue        % declaring one column version table size true
\newif\ifplainver  %declare a plain version
\ifplainver
    \confverfalse                         %automatically disable conf. version argument if it's plain
\fi

\ifconfver
      \documentclass[10pt,compsoc,journal]{IEEEtran}
\else
    \ifplainver
        \documentclass[11pt]{article}
        \usepackage{fullpage}
    \else
        \documentclass[12pt,draftcls,onecolumn]{IEEEtran}
    \fi
\fi

\usepackage{calc,amsfonts,amssymb,amsmath,bm,url,color,theorem,graphicx,cite,epstopdf,nicefrac}
\usepackage{tikz}
\usetikzlibrary{bayesnet}
\usepackage{psfrag,subfigure,float}
\usepackage[algoruled,linesnumbered]{algorithm2e}
\usepackage{multirow}
\usepackage{amssymb}% http://ctan.org/pkg/amssymb
\usepackage{pifont}% http://ctan.org/pkg/pifont
\definecolor{orange}{RGB}{255,107,0}

\definecolor{light-gray}{gray}{0.65}

\newcommand{\X}{\boldsymbol{X}}
\newcommand{\V}{\boldsymbol{V}}

\newcommand{\x}{\boldsymbol{x}}
\newcommand{\z}{\boldsymbol{z}}

\renewcommand{\u}{\boldsymbol{u}}
\renewcommand{\v}{\boldsymbol{v}}

\newcommand{\T}{{\!\top\!}}

\DeclareMathOperator*{\minimize}{\textrm{minimize}}

\newtheorem{theorem}{Theorem}
\newtheorem{lemma}{Lemma}
\newtheorem{proposition}{Proposition}

\newtheorem{remark}{Remark}

  \newcommand*\ab{.4}
\tikzset{
	net node/.style = {circle, minimum width=2*\ab cm, inner sep=0pt, outer sep=0pt, ball color=blue!50!cyan},
	net root node/.style = {net node, minimum width=3*\ab cm},
	net connect/.style = {line width=1pt, draw=blue!50!cyan!25!black},
}

\newcommand{\papertitle}{
	Link Prediction Under Imperfect Detection: Collaborative Filtering for Ecological Networks
}

\newcommand{\paperabstract}{
\textit{Matrix completion} based {\it collaborative filtering} is considered scalable and effective for online service link prediction (e.g., movie recommendation) but does not meet the challenges of link prediction in ecological networks. 
A unique challenge of ecological networks is that the observed data are subject to systematic \textit{imperfect detection}, due to the difficulty of accurate field sampling. In this work, we propose a new framework customized for ecological bipartite network link prediction. Our approach starts with incorporating the Poisson $N$-mixture model, a widely used framework in statistical ecology for modeling imperfect detection of a single species in field sampling. Despite its extensive use for single species analysis, this model has never been considered for link prediction between different species, perhaps because of the complex nature of both link prediction and $N$-mixture model inference. By judiciously combining the Poisson $N$-mixture model with a probabilistic nonnegative matrix factorization (NMF) model in latent space, we propose an intuitive statistical model for the problem of interest. We also offer a scalable and convergence-guaranteed optimization algorithm to handle the associated maximum likelihood identification problem. Experimental results on synthetic data and two real-world ecological networks data are employed to validate our proposed approach.
}

\title{\papertitle}
\author{Xiao Fu, Eugene Seo, Justin Clarke, and Rebecca A. Hutchinson
}

\begin{document}
	
	\IEEEtitleabstractindextext{%
		\begin{abstract}
			\paperabstract
		\end{abstract}
	}

	\maketitle

\section{Introduction}
{Link prediction} \cite{lu2011link} aims at inferring unseen connections between entities in a complex network, which lies at the heart of a large variety of data mining problems. For example, social network analytics involves many link prediction problems (e.g., community detection) \cite{liben2007link}. In online service recommender systems (e.g., streaming services and online shopping) \cite{ricci2015recommender,xiong2010temporal,gu2010collaborative}, link prediction plays an essential role in learning user preferences.

Among many link prediction techniques, low-rank matrix completion (MC)-based collaborative filtering (CF) \cite{yu2009fast} is one of the most popular approaches---possibly because of its simplicity and effectiveness, as well as its elegance in mathematics \cite{keshavan2010matrix,candes2010matrix}. Taking the movie recommendation problem as an example, the idea of MC-based CF is as follows: If many users have similar preferences and a lot of movies have similar traits, then the complete (but partially observed) user-movie matrix (whose entries are users' ratings for the movies) should exhibit correlations across rows and columns---and thus be an approximately low-rank matrix. Therefore, imputing those unobserved ratings should not be arbitrary---it
should be done under the constraint that the imputed movie rating matrix is low-rank. %This idea turns out to work quite well in some applications, and this seemingly simple intuition is backed by a series of mathematically rigorous works [XX-XX].

\begin{figure}
	\centering
	\includegraphics[width=0.7\linewidth]{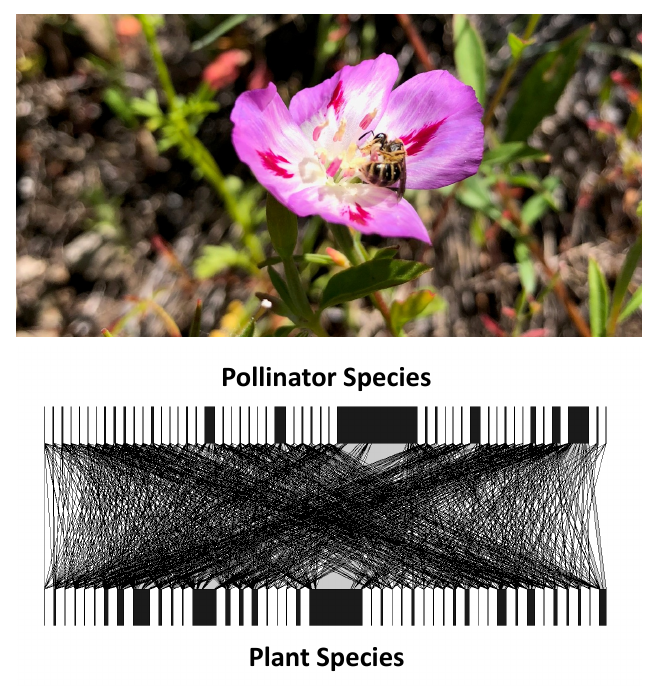}
	\caption{Dataset that motivates this work \cite{Jones2017}: Top: Halictus bee species visiting Clarkia amoena flower; 
		Bottom: Subset of the
		plant-pollinator network recorded by 
		student observers, which is subject to systematic
		under-counting.}
	\label{fig:motivation}
	\vspace{-.35cm}
\end{figure}

Existing MC-based CF approaches have been quite successful in the aforementioned online service-related domains, but note that there is a clear distinction in these applications between missing and non-missing values. For example, an online movie streaming system knows perfectly which ratings are recorded and which are missing. 
%Despite of being quite successful, the existing MC-based CF approaches seem hard to cover applications that are not closely related to online services, e.g., link prediction in ecosystems.
Our work is motivated by a domain in which this distinction is \emph{not} clear: species interaction networks. Data on these networks are often collected through field observations, which may fail to record some interactions due to limited time for sampling or poor observation conditions. 
%Note that inferring unseen links in ecosystems is very well-motivated, since many interactions between species are indeed not observed -- and thus using data mining tools to unravel such hidden connections will greatly help understand lives in our planet.
For example, hundreds of species of pollinators and plants exist in the Cascade mountains, Oregon, United States \cite{Jones2017}. Ecologists are eager to know the interaction patterns between pollinators and plants, in order to predict the effects of species extinctions or invasions and to protect ecosystem services (e.g., pollination) provided by pertinent species \cite{Tylianakis2010,Memmott2004} (see Fig.~\ref{fig:motivation} for illustration). 

At first glance, it appears that pollinator-plant link prediction is quite similar to user-movie link prediction, and that approaches proposed for the latter should also be applicable for the former. Indeed, such approaches have been attempted for these data with mixed results; CF techniques sometimes failed to outperform a baseline of simply predicting the most popular plants (i.e., items) \cite{Seo2018}. The shortcomings of traditional CF methods are due to key differences among the application domains. One particular challenge is that the data recorded for pollination networks are very different from those recorded in online services. For example, the interaction counts between pollinators and plants in the Cascade mountains were recorded by student researchers at montane meadows in the HJ Andrews Experimental Forest in Blue River, Oregon, United States; see details in \cite{Jones2017}. 
Even for meticulous observers (or even if the student observers were replaced by cameras), there is a significant chance that many interactions were not recorded for a number of reasons---since factors such as the season, location, and weather at the time of observation all play roles in the recorded number of observed interactions. This means that every entry in the pollinator-plant interaction matrix is subject to imperfect detection. For example,  ``2 interactions recorded'' may be the result of ``8 interactions really happened'' plus the effect of missed detections. Thus, the non-zero recorded counts in the interaction matrix are likely biased low. This effect applies to the recorded zeros as well; the zeros are a mixture of true zeros (no interaction) along with unseen interactions that actually occurred (missed detections). Algorithms proposed for user-movie link prediction assume that while the data may be sparse (as are pollination data), the observed ratings are essentially noise-free. However, in ecological networks, pervasive missed detections violate this assumption and demand new link prediction approaches.
%When such missed detections systematically happen across the pollinator-plant data matrix, it makes algorithms proposed for user-movie link prediction (which does not have this problem at all) hard to be applied. 

%Algorithms proposed for user-movie link prediction (which does not have this problem at all) assume that while the data may be sparse, the observed ratings are essentially noise-free. Pollination network data is also sparse, but the zeros are a mixture of true zeros (no interaction) along with unseen interactions that actually occurred (missed detections). These characteristics demand new link prediction approaches. 

%{\blue I also wanna talk something about the zeros - could be missed observation, could be real zeros. Both can be handled by Poisson modeling, and we can echo this in our `contributions' part. RAH: see sentence above? Above two paragraphs may be a bit redundant - maybe combine them on a future pass.}

The issue of imperfect detection has been recognized as a universal challenge in ecology \cite{MacKenzie2006}, since cryptic and secretive species are frequently hard to observe. A family of statistical models have been proposed to handle this problem, e.g., occupancy models for binary observations \cite{MacKenzie2002} and the $N$-mixture model for counts \cite{Royle2004}. The idea is to impose a probabilistic model that links the true species occurrence or abundance (ground-truth) and the observations (e.g., by modeling the observation as a binomial selection of the ground truth). Combined with a latent probabilistic model for the events of interest (e.g., a Poisson model), this leads to an intuitively pleasing two-layer graphical model for generating the observed data. This idea makes a lot of sense, and has been widely used in applications such as animal abundance estimation \cite{Royle2004}. 
Nevertheless, because of the intrinsic computational difficulties, it has been unclear whether this family of models could be extended to link prediction in ecological networks.
%However, this model was never introduced to ecosystem link prediction, perhaps two layer model combined with the inherent difficulties on the link prediction side leads to serious challenges in modeling, formulation, and computation.

%\smallskip

\noindent
{\bf Contributions}
In this work, we offer a computational framework for link prediction under imperfect detection.
Our work is motivated by the pollination network analysis problem mentioned before, but other ecological networks share similar properties and challenges (e.g., host-parasite networks, food webs). 
Inspired by the efficacy of statistical ecology tools for modeling imperfect detection, 
we propose a collaborative filtering framework that integrates the Poisson $N$-mixture model and low-rank nonnegative matrix factorization (NMF)---so that the challenging ecological network link prediction problem can be effectively modeled and approached.
Our detailed contributions include:

\begin{enumerate}
	\item {\bf Statistical Model for Ecological Link Prediction}
	We propose a graphical model for the generative process of observing interactions between two species in an ecosystem. Our idea is motivated by the Poisson $N$-mixture model \cite{Royle2004} that has been used for single-species observations. We impose a latent Poisson-nonnegative matrix factorization (NMF) model for generating the real counts of observations, and a Binomial-linear regression model to model the imperfect detection effect. This way, both interaction preferences (modeled by latent NMF) and factors that affect the observation (modeled by features of the regression model) can be considered under a unified framework.
	\item {\bf Effective Optimization Algorithm} We propose an effective algorithm to handle the inference problem associated with our model. The maximum likelihood (ML) estimator under the generative model poses a very challenging optimization problem, due to the Poisson distribution in the latent space, parameterized by NMF---which is NP-hard to compute. Nevertheless, we propose a \textit{block coordinate descent} (BCD)-based algorithm \cite{bertsekas1999nonlinear} to handle the ML estimator, and show that the algorithm is convergent. In addition, we judiciously design the updates so that the algorithm only consists of algebraically simple operations.
	%\item {\bf Evaluations on Pollination Network Data} We test the proposed %approach on synthetic data and a real-world pollination network prediction %problem. The pollination data are publicly available \cite{Jones2017}. %Open-sourced code for our method will be available at the authors' websites.
	\item {\bf Evaluations on Ecological Network Data} We test the proposed approach on synthetic data and two real-world ecological networks (plant-pollinator interactions and host-parasite interactions) prediction problem. The pollination and host-parasite data are publicly available; see \cite{Jones2017, Dallas2017}. Open-sourced code for our method and the baselines are also made available through GitHub at \url{https://github.com/Hutchinson-Lab/Poisson-N-mixture}.
\end{enumerate}

%\smallskip

\noindent
{\bf Notation} Throughout the paper, we will use boldface capital letters such as $\bm X$ to denote matrices. We use the Matlab notation $\bm X(i,j)$ to denote the entry of $\bm X$ at the $i$th row and $j$th column. $\bm X(i,:)$ and $\bm X(:,j)$ denote the $i$th row and the $j$th column of $\bm X$, respectively. $\bm X^\T$, $\X^{-1}$, and $\X^\dag$  denote the transpose, inverse, and pseudo-inverse of $\X$, respectively. Boldface lowercase letters such as $\bm x$ are used to denote vectors. The inequalities $\bm X\geq \bm 0$ and $\bm x\geq \bm 0$ mean that the entries of the matrix and vector are all nonnegative.

\section{Motivation and Background}
As we have mentioned briefly in the introduction,
our problem is motivated by studying pollination networks in the Cascade mountains of Oregon, United States \cite{Jones2017}.
%In this study, student researchers observed ten 3-meter by 3-meter plots in each of 12 meadows five times per summer for eight summers. They recorded the abundance of plants in each plot and all interactions observed over a 15-minute period \cite{Jones2017}.
In this study, student researchers recorded all interactions they observed between plant and pollinator species.
The goal of this study, from an ecological point of view, is to infer the structure and dynamics of the pollination network, in order to understand how the network will respond to perturbations like species extinctions and invasions. To achieve this goal, one must first infer the true pattern of connections between different species of pollinators and plants, despite the incomplete (and highly biased) data provided by field sampling. To this end, we propose an approach that combines ideas from both collaborative filtering and statistical ecology. Below, we review key concepts from these fields.

% This paragraph is already covered above, so commenting it out:
%\section{Background}
%At first glance, the problem seems to be a link prediction problem that can be handled by many existing techniques, e.g., matrix completion (MC)-based collaborative filtering (CF) that is quite successful for streaming service recommendation. If we treat pollinator-plant network as the user-movie network there, can we directly apply the existing algorithms for our problem?
%This seems plausible -- however, there are some critical differences between the two problems, as we will explain shortly.

\subsection{Matrix Completion-Based Collaborative Filtering}
One of the key techniques that enables movie recommendation is the so-called collaborative filtering (CF) \cite{zhou2008large}.
In CF, we are given a movie rating matrix $\bm X$ whose rows and columns represent users and movies, respectively---$\bm X(i,j)$ is the rating of movie $j$ by user $i$.
Only a small portion of $\X(i,j)$'s are observed, indexed by $(i,j)\in\bm \varOmega$. The task of CF is to impute or predict the missing entries, which is essentially link prediction for bipartite graphs. One of the popular formulations is
\begin{align}\label{eq:MF_CF}
\minimize_{\boldsymbol{U},\bm V}~\sum_{(i,j)\in \bm \varOmega}\left(  \X(i,j) - \bm u_i^\T\bm v_j   \right)^2
\end{align}
where $\bm u_i$ and $\bm v_j$ stand for the $i$th row and $j$th column of $\bm U\in\mathbb{R}^{I\times F}$ and $\bm V\in\mathbb{R}^{J\times F}$, respectively,
and $\bm U$ and $\bm V$ collect all the latent representations (embeddings) of the users and movies, respectively.
The rationale behind the above formulation is to model the rating of a movie by a user using the `correlation' between the user embedding and movie embedding. In addition, one can view the formulation in \eqref{eq:MF_CF} from a low-rank matrix factorization viewpoint (with many missing values)---since if many users have similar tastes and many movies have similar traits, $\bm X$ is expected to be a low-rank matrix. Hence, the collaborative filtering problem boils down to a  \textit{low-rank matrix completion} (MC) problem. 

The formulation in Eq.~\eqref{eq:MF_CF} is more suitable for continuous data. When $\bm X(i,j)$ denotes a counted number, Poisson priors are commonly used, which assumes the following observation model \cite{fevotte2009nonnegative,chi2012tensors}:
\begin{equation}
\X(i,j)\sim  {\sf Poisson}(\bm u_i^\T\bm v_j),~\bm u_i\geq \bm 0,~\bm v_j\geq \bm 0,~\forall i,j
\end{equation}        
Note that the nonnegativity constraints on $\u_i$ and $\v_j$ are needed to ensure that the Poisson parameter $\lambda_{ij}=\u_i^\T\v_j$ characterizing the expected number of interactions within a certain time interval is nonnegative.
The corresponding maximum likelihood (ML) estimator becomes
\begin{subequations}\label{eq:poissonNMF}
	\begin{align}
	\minimize_{\boldsymbol{U},\bm V}&~\sum_{i,j}\left[ \u_i^\T\bm v_j -\X(i,j)\log(\u_i^\T\bm v_j) \right]\\
	{\rm subject~to}&~\bm U\geq \bm 0,~\bm V\geq \bm 0.
	\end{align}
\end{subequations}
The problem above is essentially the same as the Kullback-Leibler (KL)-divergence based NMF, which is well-studied in the literature; see \cite{fevotte2009nonnegative,chi2012tensors}.

% Rephrased a bit 

%The MC-CF approach is well-suited for online streaming service recommendation-type link prediction but has shortcomings with respect to link prediction for ecological networks.
%In ecological networks, the number of interactions happening between two nodes in a network, e.g., pollinator $i$ and plant $j$, could be modeled by a Poisson NMF model as in \eqref{eq:poissonNMF}. However, since the observed data is subject to imperfect detection, naively applying  \eqref{eq:poissonNMF} does not account for systematic undercounting and non-randomly missing data.

Note that the Poisson modeling in \eqref{eq:poissonNMF} has several appealing features. Most notably, it naturally models missing elements in $\bm X$, since $\X(i,j)=0$ is allowed to happen with positive probability under a Poisson distribution. 
This way, the model interprets two major reasons for having missing links (namely, interactions do not exist and interactions were not observed) in a ecological network using a simple and unified way.
%{\blue This way, the model interprets two major reasons for having missing links (namely, interactions do not exist and interactions were not observed) in a ecological network using a simple and unified way.} 
The Poisson NMF model is preferred also because many ecological data are recorded as counts of interactions or encounters---which are both natural numbers rather than continuous values.
This makes the formulation in \eqref{eq:poissonNMF} more appealing for link prediction in ecosystems. Nevertheless, the Poisson NMF model does not consider the imperfect detection effect, and thus naively applying  \eqref{eq:poissonNMF} does not account for systematically under-counted data. In particular, the mixture of both true and false zeros leads to over-dispersion (zero-inflation) of the Poisson distribution if imperfect detection is not incorporated.
%{\blue I'm concerned that this paragraph may be a little inconsistent in talking about the zeros. First we say that it can interpret two kinds of 0s in a simple, unified way. then we say that having true and false zeros leads to overdispersion. I suggest we delete the blue sentence above.}

\subsection{Poisson $N$-Mixture Model}
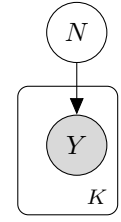
\begin{figure}[t]
	\centering
	\begin{tikzpicture}[]
\node at (   0,1.5) [circle,draw,minimum size=8mm,inner sep=-5pt] (Xt) {$N$};
%\node at (-1.5,1.5) [circle,draw,minimum size=8mm,inner sep=-5pt] (Xt-1) {$\lambda$};
%\node at ( 1.5,1.5) [circle,draw,minimum size=8mm,inner sep=-5pt] (Xt+1) {$lambda$};
\node at (   0,0) [circle,draw,minimum size=8mm,inner sep=-5pt,fill=gray!30] (Yt) {$Y$};
%\node at (-1.5,0) [circle,draw,minimum size=8mm,inner sep=-5pt] (Yt-1) {$p$};
%\node at ( 1.5,0) [circle,draw,minimum size=8mm,inner sep=-5pt,fill=gray!30] (Yt+1) {$Y_{t+1}$};
%\node at (-2.9,1.5) (before) {\dots};
%\node at ( 2.9,1.5) (after)  {\dots};
%\draw [->] (before) -- (Xt-1);
%\draw [->] (Xt+1) -- (after);
%\draw [->] (Xt-1) -- (Xt);
%\draw [->] (Xt) -- (Xt+1);
%\draw [->] (Yt-1) -- (Yt);
\draw [->] (Xt) -- (Yt);
\plate[inner sep=0.25cm, xshift=-0.12cm, yshift=0.12cm] {plate1} {(Yt)} {$K$};
%\draw [->] (Xt+1) -- (Yt+1);
\end{tikzpicture}
	\caption{Graphical model of a Poisson $N$-mixture model when $K$ observations are made for a single spices.}
	\label{fig:hmm}
	\vspace{-.35cm}
	%	\vspace{-.75cm}
\end{figure}

In statistical ecology, the Poisson $N$-mixture model is applied widely to handle the imperfect detection effect in field sampling---especially when observing animals. The model is simple and intuitive: the true abundance of an animal is modeled as \[N\sim{\sf Poisson}(\lambda).\] The observed count is modeled as a binomial selection of $N$, i.e.,
\[        Y \sim {\sf Binominal}(N,p)    \]
where $p$ is the detection probability; see the illustration in Fig.~\ref{fig:hmm}.
This model admits inference on both the true and observed counts by explicitly accounting for the observation process. Both $\lambda$ and $p$ can be linked to features that affect true abundance (e.g., habitat characteristics) and detection probability (e.g., weather conditions), respectively.
However, most applications of this model have been concerned with estimating $p$ and $\lambda$ for \textit{single species} \cite{Royle2004,dennis2015computational}, and this model has not been applied to species interaction networks. The standard $N$-mixture model is already computationally hard \cite{dennis2015computational}, so incorporating it into link prediction problems is quite challenging. %We offer a solution to this problem in this work.

\section{Proposed Approach}
Our approach combines the strengths of the Poisson $N$-mixture model for estimating species abundance under imperfect detection and low-rank CF models for link prediction. We propose the following generative model:
\begin{subequations}\label{eq:generative}
	\begin{align}
	&\lambda_{ij} =  \bm u_i^\T\bm v_j,~\bm u_i\geq\bm 0,~\bm v_j\geq\bm 0, \\
	&N_{ij} \sim {\sf Poisson}(\lambda_{ij}) \\
	&p_{ij} =\bm  \alpha^\T \z^{(ij)},\quad 0\leq p_{ij}\leq 1\\
	& y_{ij} \sim {\sf Binomial}(N_{ij}, p_{ij}) 
	\end{align}
\end{subequations}
We model $\lambda_{ij}$, i.e., the Poisson parameter for the $(i,j)$th interaction. 
A bigger $\lambda_{ij}$ means that the interaction occurs more frequently on average.
We assume that the interaction between pollinator $i$ and plant $j$ is more frequent if their embeddings are more `correlated'. %{\blue Why do we keep putting correlated in quotes? There's another spot above. Do we mean something other than true correlation?} 
Note that we constrain $\boldsymbol{U}$ and $\bm V$ to be elementwise nonnegative, because $\lambda_{ij}$ has to be nonnegative.

In the observation model, we let the detection probability of the $(i,j)$th interaction be $p_{ij} =\bm  \alpha^\T \z^{(ij)}$, where ${\bm z}^{(ij)}\in\mathbb{R}^R$ is a interaction-specific feature vector. This vector is defined over a feature space that affects the detection probability for a particular pair of $(i,j)$---e.g., abundance of plant $j$, size of pollinator $i$, and the time of observation can all be features collected by $\z^{(ij)}$. This way, side information that we may have in the data generating process can be seamlessly incorporated into the model. %---a side benefit of the presumed generative model in \eqref{eq:generative}. 
Note that $p_{ij}$ is a detection probability, and thus naturally lies in between 0 and 1.

\section{Optimization Algorithm}
Following the generative model in \eqref{eq:generative}, we are interested in learning $\boldsymbol{U}$, $\bm V$, and $\bm \alpha$.
Let us denote the probability mass function of $Y_{ij}$ by 
\[     {\rm Pr}(Y_{ij}=y_{ij};\lambda_{ij},p_{ij}).  \]
By the assumed generative model and the law of total probability, the log-likelihood function of the parameters of interest is as follows:
%\begin{subequations}\
\begin{align*}\label{eq:prob_form}
& \log\left(\prod_{ij} \sum_{n=y_{ij}}^\infty {\sf Pr}(N_{ij}=n;\lambda_{ij}) {\sf Pr}(y_{ij}|N_{ij}=n; p_{ij})\right) \\
= &\sum_{ij} \log\left( \sum_{n=y_{ij}}^\infty\left(\frac{\lambda_{ij}^{n} e^{-\lambda_{ij}}}{n!} \frac{n!}{y_{ij}!(n-y_{ij})!} p_{ij}^{y_{ij}} (1-p_{ij})^{n-y_{ij}}\right)\right) 
\end{align*}
%\end{subequations}
%The infinite sum here is very hard to handle, in terms of optimization. To handle it, consider the following derivation:
%\begin{align}
%&\sum_{n=y_{ij}}^\infty \frac{\lambda_{ij}^{n} e^{-\lambda_{ij}}}{n!}  \frac{n!}{y_{ij}!(n-y_{ij})!} p^{y_{ij}} (1-p)^{n-y_{ij}}\\
%& = \frac{p^{y_{ij}}e^{-\lambda_{ij}}}{y_{ij}!} \sum_{n=y_{ij}}^\infty \frac{\lambda_{ij}^{n}(1-p)^{n-y_{ij}}}{(n-y_{ij})!}.
%\end{align}
%Now, let us do the following change of variables
%\[  m_{ij} = n-y_{ij}.  \]
%Then, what we have is
%\begin{align}
%&\sum_{n=y_{ij}}^\infty \frac{\lambda_{ij}^{n} e^{-\lambda_{ij}}}{n!}  \frac{n!}{y_{ij}!(n-y_{ij})!} p^{y_{ij}} (1-p)^{n-y_{ij}}\\
%& = \frac{p^{y_{ij}}e^{-\lambda_{ij}}}{y_{ij}!} \sum_{n=y_{ij}}^\infty \frac{\lambda_{ij}^{n}(1-p)^{n-y_{ij}}}{(n-y_{ij})!}\\
%& = \frac{p^{y_{ij}}e^{-\lambda_{ij}}}{y_{ij}!} \sum_{m_{ij=0}}^\infty \frac{\lambda_{ij}^{m_{ij}+y_{ij}}(1-p)^{m_{ij}}}{m_{ij}!}\\
%& = \frac{(p\lambda_{ij})^{y_{ij}}e^{-\lambda_{ij}}}{y_{ij}!} \sum_{m_{ij=0}}^\infty \frac{\lambda_{ij}^{m_{ij}}(1-p)^{m_{ij}}}{m_{ij}!}\\
%& =  \frac{(p\lambda_{ij})^{y_{ij}}e^{-\lambda_{ij}p}}{y_{ij}!} %\sum_{m_{ij=0}}^\infty \frac{\lambda_{ij}^{m_{ij}}(1-p)^{m_{ij}}}{m_{ij}!}
%\end{align}
%where the last equality is by the Taylor expansion. Hence, the ML estimator becomes
The above formulation has an infinite sum, which appears problematic at first glance:
If one directly applies maximum likelihood estimation using the above, then an infinite number of terms are involved in the estimator,
which might be very hard to handle.

To proceed, we invoke the following lemma \cite{dennis2015computational}:
\begin{lemma}\label{lem:key}
	The following equality holds:
	\[\sum_{n=y_{ij}}^\infty \frac{\lambda_{ij}^{n} e^{-\lambda_{ij}}}{n!}  \frac{n!}{y_{ij}!(n-y_{ij})!} p^{y_{ij}} (1-p)^{n-y_{ij}}=  \frac{(p\lambda_{ij})^{y_{ij}}e^{-\lambda_{ij}p}}{y_{ij}!}. 	\]
\end{lemma}

Lemma~\ref{lem:key} is a classic result that is employed in many fields \cite{dennis2015computational}---to be self-containing, we present a short proof in Appendix~\ref{app:infSumDeriv}.
Nonetheless, the result is very useful in our context, since it elegantly converts a quite complicated ML estimator to one that is much easier to maneuverer for subsequent steps.

Applying Lemma~\ref{lem:key}, we have the following simplified log-likelihood:
\begin{equation}\label{eq:simplified}
\begin{aligned}
&\sum_{ij} \log\left( \frac{\left(p_{ij}\lambda_{ij}\right)^{y_{ij}}e^{-\lambda_{ij}p_{ij}}}{y_{ij}!}\right) \\
&=  \sum_{ij}\left( y_{ij}\log p_{ij} +y_{ij}\log \lambda_{ij} - \lambda_{ij}p_{ij} -\log y_{ij}!\right).
\end{aligned}
\end{equation}
%The derivation from (XX) to (XX) is put in Appendix XX for being self-containing. 
%Note that the optimization problem is still very challenging; i.e., we aim to optimize
Hence, under our model, we aim to optimize the following: 
\begin{subequations}\label{eq:ml}
	\begin{align}
	\minimize_{\boldsymbol{U},\V,\bm \alpha}~&-\sum_{ij}\left[ y_{ij}\log\left(\bm \alpha^\T\bm z^{(ij)}\right) +y_{ij}\log \u_i^\T\v_j \right. \nonumber\\
	&\quad\quad\quad \left.- (\u_i^\T\v_j)\left(\bm \alpha^\T\bm z^{(ij)}\right)  \right] \nonumber	\\
	{\rm subject~to:}~&\u_i\geq\bm 0,~\v_j\geq \bm 0,~\forall i,j\\
	&0\leq \bm \alpha^\T\bm z^{(ij)} \leq 1,~\forall i,j.
	\end{align}
\end{subequations}
The problem is still very challenging because of the coupled nature of $\boldsymbol{U},\V$, and $\bm \alpha$. In addition, the constraints on the optimization variables are not straightforward to handle. 
To tackle the formulated maximum likelihood estimation problem,
we propose a {\it block coordinate descent} (BCD) \cite{bertsekas1999nonlinear} based algorithm.
The basic idea is to cyclically solve `partial' optimization problems w.r.t. a single block variable among $\bm U$, $\V$, and $\bm \alpha$, respectively, while fixing the other two.
This way, the subproblems can be handled efficiently---leveraging computational tools for probabilistic NMF.
The high-level algorithmic structure is summarized in Fig.~\ref{algo:BCD}, where we use $f(\bm U,\bm V, \bm \alpha)$ and $t$ to denote the objective function in \eqref{eq:ml} and the index of iteration, respectively. The updates are carried out in each iteration until a certain convergence criterion is met.

\begin{figure}
	\centering
	\boxed{
		\begin{minipage}{.8\linewidth}\		repeat the following until convergence:
			\begin{align*}
			\bm \alpha^{t+1}&\leftarrow \arg\min_{ 0\leq \bm \alpha^\T\bm z_{ij} \leq 1,~\forall i,j}~f(\bm U^t,\bm V^t, \bm \alpha)\\
			\bm U^{t+1}&\leftarrow \arg\min_{\bm U\geq \bm 0}~f(\bm U,\bm V^t, \bm \alpha^{t+1})\\
			\bm V^{t+1}&\leftarrow \arg\min_{\bm V\geq \bm 0}~f(\bm U^{t+1},\bm V, \bm \alpha^{t+1})\\
			t&\leftarrow t+1
			\end{align*}
	\end{minipage}}
	\caption{The algorithmic framework for the ML estimator in Eq.~\eqref{eq:ml}.}\label{algo:BCD}
	\vspace{-.35cm}
\end{figure}

\subsection{The $\alpha$-update}
First, consider the $\bm \alpha$-update.
Since $\bm U,\bm V$ are fixed, the relevant part of the objective function \eqref{eq:ml} is:
\begin{equation}\label{eq:alpha_sub}
\begin{aligned}
\minimize_{\bm \alpha}&~\sum_{(i,j)\in\Omega} -\left[  y_{ij}\log \left( \bm \alpha^\T\bm z^{(ij)} \right)  - \left( \bm \alpha^\T \bm z^{(ij)} \right) \lambda_{ij} \right]\\
{\rm s.t.}&~ 0\leq \bm \alpha^\T\bm z^{(ij)} \leq 1,
\end{aligned}
\end{equation}
where we have omitted the superscript `$t$' for notational simplicity.
We use $\lambda_{ij}=\u_i^\T\v_j$ to denote the current estimate of $\lambda_{ij}$.
This is a convex optimization problem. Nevertheless, there is a serious scalability issue if standard optimization toolboxes (e.g., CVX and the interior point method (IPM) \cite{CVX,grant2008cvx}) are employed. The scalability challenge lies in the large number of constraints, i.e., $IJ$ constraints, which makes IPM-like algorithms very slow (i.e., $O(R^2IJ)$ flops are needed and both $I$ and $J$ can be large). 

To circumvent the scalability issue, we propose an {\it alternating direction method of multipliers} (ADMM) \cite{Boyd11}  based algorithm. As will be seen, with judicious reformulation,
the updates of the ADMM algorithm are all very lightweight.
Specifically, we introduce an auxiliary variable $\bm P\in\mathbb{R}^{I\times J}$ such that $\bm P(i,j)=p_{ij}=\bm \alpha^\T\bm z^{(ij)}$ and we denote
\[ \bm p = {\rm vec}(\bm P). \] 
Consequently, Problem~\eqref{eq:alpha_sub} can be rewritten as 
\begin{equation}
\begin{aligned}
\min_{\bm \alpha,\bm p}&~\sum_{ij} -\left[  y_{ij}\log  p_{ij} -  p_{ij}\lambda_{ij} \right]\\
{\rm s.t.}&~ 0\leq p_{ij} \leq 1,~\bm \alpha^\T\bm z^{(ij)}=p_{ij}.
\end{aligned}
\end{equation}
The augmented Lagrangian is
\[ {\mathcal L} = \sum_{ij} -\left[  y_{ij}\log  p_{ij} -  p_{ij}\lambda_{ij} \right] +  \frac{\rho}{2}\|{\bm p}-\bm Z\bm \alpha +\bm \omega\|_2^2   \]
where $\bm p$ is the vectorized version for $\bm P$, $\rho>0$ is a pre-specified regularization parameter (see details in \cite[Chapter 3]{Boyd11} for how to select this parameter), $\bm \omega={\rm vec}(\bm W)$, $\bm W(i,j)$ is the dual variable associated with $\bm \alpha^\T\bm z^{(ij)}=p_{ij}$, and
\[  \bm Z \in\mathbb{R}^{IJ\times R},~\bm Z((j-1)I+i,:)=(\bm z^{(ij)})^\T,\]
which is a matrix that collects all $(\bm z^{(ij)})^\T$'s as its rows.
The updates of the ADMM algorithm are as below:
\begin{subequations}\label{eq:admm}
	\begin{align}
	\bm p &\leftarrow \arg\min_{{\bm 0}\leq\bm p\leq {\bm 1}}~\sum_{ij} -\left[  y_{ij}\log  p_{ij} -  p_{ij}\lambda_{ij} \right] \nonumber \\
	&\quad\quad\quad\quad\quad\quad\quad\quad  +  \frac{\rho}{2}\|{\bm p}-\bm Z\bm \alpha +\bm \omega\|_2^2 \label{eq:psub}\\
	{\bm \alpha}& \leftarrow \arg\min_{\bm \alpha}~\|{\bm p}-\bm Z\bm \alpha +\bm \omega\|_2^2 \label{eq:asub} \\
	{\bm \omega}&\leftarrow {\bm p}-\bm Z\bm \alpha +\bm \omega
	\end{align}
\end{subequations}
In \eqref{eq:admm}, the dual update (i.e., $\bm \omega$) is trivial and naturally lightweight.
The second line can be updated via the following closed form solution, since it is simply a least squares problem:
\begin{equation}
\bm \alpha \leftarrow \bm Z^\dag (\bm p+\bm \omega),
\end{equation}
and the pseudo-inverse $\bm Z^\dag$ can be computed and stored in advance (which only needs to be computed once before the whole BCD algorithm starts)---meaning that the only operation needed here is matrix-vector multiplication.

It seems that the most difficult subproblem is Problem~\eqref{eq:psub}. Nonetheless, this subproblem can be shown to admit a closed-form solution:
\begin{lemma}\label{lem:p}
	The solution to Problem~\eqref{eq:psub} can be obtained by the following algebraic form:
	\begin{align}\label{eq:psol}
	&\bm p  \leftarrow 
	\left[\frac{(\rho \bar{\bm p} -\bm \lambda) + \sqrt{(\rho \bar{\bm p}-\bm \lambda)^2+4\rho \bm y}}{2\rho}\right]_{[0,1]},
	\end{align}
	where all the power, square, and division operators are taken elementwise, $[\bm Q]_{[0,1]} = \min[\max[\bm Q,{\bm 0}],{\bm 1}]$, and
	\[\bm y= {\rm vec}(\tilde{\bm Y}). \]
\end{lemma}

{\it Proof}:
Notice that the problem is separable w.r.t. each $(i,j)$, and thus it suffice to show \eqref{eq:psol} holds for each $p_{ij}$.
In other words, Problem~\eqref{eq:psub} can be solved via solving the following scalar problem:
\[  \minimize_{ 0\leq p\leq 1}~ -  y\log  p +  p\lambda +  \frac{\rho}{2}(p-\bar{p})^2 \]
where we omitted the subscripts and let
\[  \bar{\bm p} = \bm Z\bm \alpha -\bm \omega. \]

First consider the case where $y>0$. Then, the optimal solution must satisfy $p>0$, since $p=0$ results in an infinitely large objective value.
%Since we have a scalar function $f(p)$, living in $p\in[0,1]$ {\blue not a sentence}. 
This univariate function in the objective is \textit{convex,} and thus can be easily solved via taking the first-order derivative w.r.t. $p$ and setting it to zero, which leads to
\[    \frac{-y}{p} + \lambda + \rho(p-\bar{p})=0.                                \]
The above amounts to a root-finding problem for a second-order polynomial.
Hence, the optimal solution is either at the boundary of $p\in[0,1]$ (in this case, only $1$) or
\begin{align}\label{eq:psoln}
&p  =
\left[\frac{(\rho \bar{p} - \lambda) + \sqrt{(\rho \bar{p}- \lambda)^2+4\rho  y}}{2\rho}\right],
\end{align}
if the above is smaller than 1.

Also consider the case where $y=0$. This is even simpler since the problem becomes
\[  \minimize_{ 0\leq p\leq 1}~ +  p\lambda +  \frac{\rho}{2}(p-\bar{p})^2, \]
which is a univariate quadratic problem. 
One can show that
\[  p =\frac{(\rho \bar{p} -\lambda) + \sqrt{(\rho \bar{p}-\lambda)^2}}{2\rho} \]
or on the boundary of $[0,1]$ by the same reasoning.
Therefore, the solution for $\bm p$ can be summarized as that in \eqref{eq:psol}.
%\begin{align}\label{eq:psoln}
%&\bm p  \leftarrow 
%\left[\frac{(\rho \bar{\bm p} -\bm \lambda) + \sqrt{(\rho \bar{\bm p}-\bm \lambda)^2+4\rho \bm y}}{2\rho}\right]_{[0,1]},
%\end{align}
%This completes the proof. 
\hfill $\square$

%in which
%\begin{align*}
%\tilde{\bm Y}(i,j) =\begin{cases}
%y_{ij}~&~(i,j)\in\Omega\\
%0,~&~{\rm o.w.}
%\end{cases}
%\end{align*}

The ADMM algorithm is summarized in Fig.~\ref{algo:admm}. One can see that all the updates are very lightweight and thus the algorithm is easily scalable.
Since the $\bm \alpha$-subproblem is convex, the ADMM algorithm guarantees to solve it to optimality \cite{Boyd11}.
\begin{figure}
	\centering
	\boxed{
		\begin{minipage}{.8\linewidth}
			repeat the following until convergence:
			\begin{align*}
			&\text{update $\bm p\leftarrow$ Eq~\eqref{eq:psol};}\\
			&\text{update $\bm \alpha\leftarrow \bm Z^\dag(\bm \omega +\bm p)$;}\\
			&\text{update $\bm \omega\leftarrow {\bm p}-\bm Z\bm \alpha +\bm \omega$;}
			\end{align*}
	\end{minipage}}
	\caption{The ADMM algorithm for solving \eqref{eq:alpha_sub}.}\label{algo:admm}
	\vspace{-.35cm}
\end{figure}

\subsection{The $U$-update}
The subproblem w.r.t. $\bm u_i$ and $\bm v_j$ is
\begin{equation}\label{eq:kl_nmf}
\begin{aligned}
\min_{{\bm u}_i\geq 0,{\bm v}_j\geq {\bm 0}} \sum_{i,j}\left[p_{ij}{\bm u}_i^\T\bm v_j - y_{ij}\log ({\bm u}_i^\T\bm v_j) \right],
\end{aligned}
\end{equation}
where $p_{ij}=\bm \alpha^\T\bm z^{(ij)}$.
In essence, the above problem can be understood as a weighted version of KL-divergence based nonnegative matrix factorization--- the only difference between \eqref{eq:kl_nmf} and KL-divergence NMF is that the terms $\u_i^\T\v_j$ are scaled by $p_{ij}$.
Techniques such as multiplicative updates (MU) \cite{seung2001algorithms} that are used for KL-divergence NMF can still be applied here, with careful modifications.
To be specific, consider  the objective function of the $\boldsymbol{U}$-subproblem,
which can be decoupled to $I$ subproblems w.r.t. $\u_i$ as follows:
\[  f(\bm u_i) = \sum_{j} \left[p_{ij}{\bm u}_i^\T\bm v_j - y_{ij}\log ({\bm u}_i^\T\bm v_j) \right] \]
and its tight upper bound
\begin{align}\label{eq:u_tilde}
g(\bm u_i,\bar{\bm u}_i) =    &{\bm u}_i^\T \underbrace{\left(\sum_{j}p_{ij}\bm v_j \right) }_{\tilde{\bm u}_i} - \sum_{j} y_{ij}\sum_{r=1}^R\beta_r^{(ij)}\log\left( \frac{u_{i,r} v_{j,r}}{\beta_r^{(ij)}} \right), 
\end{align} 
where
$  \beta_r^{(ij)} = \frac{\bar{u}_{i,r}v_{j,r}}{\bar{\bm u}_i^\T\bm v_j} .$
Note that the upper bound is derived from the Jensen's inequality; see details in \cite{chi2012tensors,fevotte2009nonnegative}.
The upper bound is tight in the following sense:
\begin{subequations}
	\begin{align}\label{eq:upp}
	&f(\bm u_i) \leq  g(\bm u_i,\bar{\bm u}_i) ,~\forall \bm u_i\geq \bm 0,\\
	&f(\bar{{\bm u}_i}) =  g(\bar{\bm u_i},\bar{\bm u}_i)\\
	&\nabla_{\bm u_i}f(\bar{{\bm u}_i}) =  \nabla_{\bm u_i}g(\bar{\bm u_i},\bar{\bm u}_i);
	\end{align}
\end{subequations}
i.e., the two functions are `tangent' at $\bm u_i=\bar{\bm u}_i$.

The idea of MU-like algorithms is to update $\u_i$ via solving
\begin{equation}\label{eq:g_sub}
\u_i \leftarrow  \arg\min_{\u_i\geq \bm 0}~g(\bm u_i,\bar{\bm u}_i),
\end{equation}
instead of directly solving $f(\u_i)$ that is hard to tackle. This is widely known to be {\it majorization minimization} (MM) in the optimization literature \cite{razaviyayn2013unified}, which ensures decreasing the cost value of $f(\u_i)$ in each iteration.

To solve the surrogate optimization problem \eqref{eq:g_sub}, let us take derivative w.r.t. $u_{i,r}$, we have
\begin{align*}
\nabla_{u_{i,r}}g(\bm u_i,\bar{\bm u}_i) &= \tilde{u}_{i,r} -\sum_{j} y_{i,j} \beta_{r}^{(ij)}\left( \frac{\beta^{(ij)}_r}{u_{i,r}v_{j,r}} \right)\left(\frac{ v_{j,r}}{\beta_r^{(ij)} } \right)\\
& =  \tilde{u}_{i,r} - \left(\sum_{j} y_{i,j} \beta_{r}^{(ij)}\right) \frac{1}{u_{i,r}}.
\end{align*}
Setting the above equal to zero leads to
\begin{equation}
u_{i,r} =  \frac{\left(\sum_{j} y_{i,j} \beta_{r}^{(ij)}\right)}{\tilde{u}_{i,r}},
\end{equation}
which is surely nonnegative, being reminiscent of the multiplicative update (MU)---under the condition that $\tilde{u}_{i,r}$ is nonzero, which hardly happens in practice if $\bm U$, $\V$ and $\bm P$ are initialized with nonzero matrices.

By role symmetry, we immediately have the following:
\begin{equation}\label{eq:uri}
v_{j,r} =  \frac{\left(\sum_{i} y_{i,j} \beta_{r}^{(ij)}\right)}{\tilde{v}_{j,r}},\quad \tilde{\bm v}_{j} = \sum_{i} {p_{i,j}\bm u_i}
\end{equation}

The update of $\bm U$ and $\bm V$ can be represented in a more compact form, i.e.,
\begin{align}\label{eq:UV}
&\bm U \leftarrow  \left(\bm U \circ \bm \Phi\right) / \tilde{\bm U},~\bm \Phi =\left(\bm Y/\bm U\bm V^\T \right)\bm V\\
&\V \leftarrow \left(\bm V \circ \bm \Psi \right) / \tilde{\bm V},~\bm \Psi =\left(\bm Y^\T/\bm V\bm U^\T \right)\bm U,
\end{align}
where ``$\circ$'' denotes the Hadamard product and ``$/$'' denotes the elementwise division (i.e., ``$./$'' in {\sf Matlab}).
One can see that the above updates indeed exhibit a flavor of MU, with the newly defined $\tilde{\bm U}$ and $\tilde{\bm V}$ caused by incorporating the weighting matrix $\bm P$; see Eqs~\eqref{eq:u_tilde} and \eqref{eq:uri} for the definitions of the $i$th row of $\tilde{\bm U}$ and $j$th row of $\tilde{\bm V}$, respectively.

\begin{remark}\label{rmk:imputation}
	Our formulation and algorithm naturally work for networks with many zeros without knowing if they are ``undetected links'' (observation made but no interaction between the associated pairs detected) or ``true misses'' (no observation made for the associated pairs).
	One remark is that in many cases, domain knowledge can be used to distinguish ``undetected links'' and ``true misses'', at least for part of the network.
	In incorporating such domain knowledge is often useful, especially when dealing with noisy real-world data.
	Minor modifications to the algorithm suffice to take ``true misses'' into consideration. To be specific, we can keep the same $\bm \alpha$-update as before, but only applied to the entries where observations were made (no matter if interactions were recorded). For the $\bm U,\bm V$-updates, we first impute the missing entries (where no observations were made) $\widehat{y}_{i,j}=\widehat{\bm u}_i^\T\widehat{\bm v}_j$, where $\widehat{\bm u}_i$, $\widehat{\bm v}_j$ denote the current estimations for $\bm u_i$ and $\bm v_j$, respectively, and then apply the proposed algorithm. This simple heuristic admits an \textit{expectation-maximization} (EM) interpretation in the NMF literature \cite{kim2009weighted}.
\end{remark}

\subsection{Convergence Properties}
Our algorithm involves an exactly solved block subproblem (i.e., the $\bm \alpha$) and two inexactly solved blocks (i.e., the $\bm U$ and $\V$ blocks). One natural question is: does the algorithm converge? The answer is affirmative, under certain conditions. This can be shown by invoking recent results regarding \textit{inexact} block coordinate descent \cite{razaviyayn2013unified,hong2016unified}. Specifically, we have the following proposition:
\begin{proposition}\label{prop:convergence}
	Assume that the $\bm \alpha$-subproblem is solved by the ADMM algorithm in each iteration, and $\bm U$ and $\bm V$ are updated by MU in each iteration. Also assume that there is no zero element appearing in $\bm U$ and $\bm V$ throughout the iterations. Then, every limit point of the solution sequence produced by the proposed algorithm is a stationary point of Problem~\eqref{eq:ml}.
\end{proposition}
Proposition~\ref{prop:convergence} asserts that the algorithm is convergent under certain conditions---and every limit point satisfies the necessary conditions for optimality.
%The proof leverages the recent advances in nonlinear programming, in particular, the \textit{ successive upper bound minimization} (BSUM) framework \cite{razaviyayn2013unified}.
The detailed proof is relegated to Appendix~\ref{app:converge}.
Since we use a MU-like algorithm for updating $\bm U$ and $\bm V$, the caveat in theory is that convergence is only guaranteed if there is no zero elements appeared in $\bm U$ and $\bm V$ in each iteration---which is not easy to check or ensure. 
Even worse, if zero appears in $\bm U$ or $\bm V$, the update rules such as Eq.~\eqref{eq:uri} are ill-defined.
Nevertheless, there are many pragmatic ways to `robustify' the algorithm, e.g., by adding a very small $\epsilon>0$ to $\tilde{u}_{i,r}$ and $\tilde{v}_{j,r}$ in each iteration \cite{chi2012tensors} [cf. Eq.~\eqref{eq:uri}]. This usually improves the performance of MU-type algorithms \cite{chi2012tensors,seung2001algorithms} (also see Appendix~\ref{app:converge}).

%\subsection{Handing Certain and Uncertain Misses}
%Our basic algorithm considers the most general case for ecological networks, where one normally does not have prior knowledge about if an entry is unobserved or zero---i.e., all the zeros are ``uncertain misses''.
%Nonetheless, we notice there might be cases where we are certain about the zeros. For example, it is possible that by domain knowledge, we know that there are some species in the network who never interact with each other. 
%Incorporating such domain knowledge or prior information may enhance performance.
%
%To this end, a simple extension to our original formulation in \eqref{eq:ml} can be considered:
%\begin{subequations}\label{eq:ml_misses}
%	\begin{align}
%	\minimize_{\boldsymbol{U},\V,\bm \alpha}~&-\sum_{(i,j)\in {\bm \Omega}}\left[ y_{ij}\log\left(\bm \alpha^\T\bm z^{(ij)}\right) +y_{ij}\log \u_i^\T\v_j \right. \nonumber\\
%	&\quad\quad\quad \left.- (\u_i^\T\v_j)\left(\bm \alpha^\T\bm z^{(ij)}\right)  \right] \nonumber	\\
%	{\rm subject~to:}~&\u_i\geq\bm 0,~\v_j\geq \bm 0,~\forall i,j\\
%	&0\leq \bm \alpha^\T\bm z^{(ij)} \leq 1,~\forall i,j.
%	\end{align}
%\end{subequations}
\section{Numerical Experiments}
In this section, we offer a series of numerical results to showcase the effectiveness of the proposed algorithm.
We start with synthetic data that obeys the considered generative model, as a sanity check for convergence, complexity, and estimation accuracy.
Then, we use the pollination network data collected in the Cascades in Oregon \cite{Jones2017} and the host-parasite network data collected in a New Mexican desert ecosystem \cite{dallas2017predicting} to evaluate the performance of the proposed approach on real ecological network link prediction problems.

\subsection{Synthetic Data}
We generate the synthetic data following the model described in \eqref{eq:generative}. Specifically, we first generate $\boldsymbol{U}$ and $\bm V$ whose elements are drawn uniformly at random between zero and $\gamma>0$---note that we change $\gamma$  under different problem settings such that $\bm u_i^\T\bm v_j$ is not too small or too large, leading to pathological cases. We also enforce the \textit{separability condition} on $\bm U$ and $\bm V$ so that the underlying NMF model is identifiable \cite{fu2018nonnegative,donoho2003does,huang2014non}. Then, we generate the feature vectors $\bm z^{(ij)}$ and the linear regression coefficients $\bm \alpha$ following the same uniform distribution.

For synthetic data, we use two baselines as benchmarks. The first one is the Poisson NMF algorithm \cite{chi2012tensors,seung2001algorithms} that handles Problem~\eqref{eq:poissonNMF}, without considering Binomial selection. The second algorithm is MC-based CF that handles a similar problem as in \eqref{eq:MF_CF}, with regularizations for enhancing performance \cite{koren2009matrix}.

We adopt a straightforward evaluation strategy for simulated data. Since we know the ground-truth model parameters $\bm U$, $\bm V$, and $\bm \alpha$, we use mean squared error (MSE) to evaluate performance; e.g., the MSE of the estimated $\bm U$ is defined as following:
\begin{align*}
&{\sf MSE}=\min_{\pi(f)\in\{1,\ldots,F\}}\frac{1}{F}\sum_{f=1}^F\left\| \frac{\bm U(:,{\pi(f)})}{\|\bm U(:,{\pi(f)})\|_2} -\frac{\widehat{\bm U}(:,f)}{\|\widehat{\bm U}(:,f)\|_2}\right\|_2^2
\end{align*}
where $\widehat{\bm U}$ denotes the estimate of $\bm U$ and $\pi(f)$'s are under the constraint $\{\pi(1),\ldots,\pi(F)\}=\{1,\ldots,F\}$---i.e., $[\bm U(:,\pi(1)),\ldots,\bm U(:,\pi(F))]$ is a column-permuted version of $\bm U$. We use the above because NMF has intrinsic scaling and permutation ambiguities, which need to be fixed before evaluation \cite{fu2018nonnegative,gillis2014and}. For $\bm V$ we use the same evaluation. For $\bm \alpha$, we measure the MSE by
${\sf MSE} = \frac{1}{R}\left\| \widehat{\bm \alpha} - \bm \alpha \right\|_2^2,$
where $\widehat{\bm \alpha}$ denotes the algorithm-estimated $\bm \alpha$.

All the algorithms under test in this subsection are coded in Matlab, and are all initialized by the same random $\bm U$ and $\bm V$. For the proposed method, $\bm \alpha$ is also randomly initialized.
All the entries of the initializations of $\bm U,\bm V$, and $\bm \alpha$ are drawn uniformly at random between zero and one.

%\begin{figure}
%	\centering
%	\includegraphics[width=0.7\linewidth]{figures/itervsacc}
%	\caption{The average MSEs of the estimated $\bm U,\bm V$ produced by the algorithms under test.}
%	\label{fig:itervsacc}
%\end{figure}
%
%
%\begin{figure}
%	\centering
%	\includegraphics[width=0.7\linewidth]{figures/itervsalphaacc}
%	\caption{The average MSE of the estimated $\bm \alpha$ produced by the proposed algorithm.}
%	\label{fig:itervsalphaacc}
%\end{figure}

\begin{figure}
	\centering
%	\begin{minipage}{.5\linewidth}
		\centering
		\includegraphics[width=.7\linewidth]{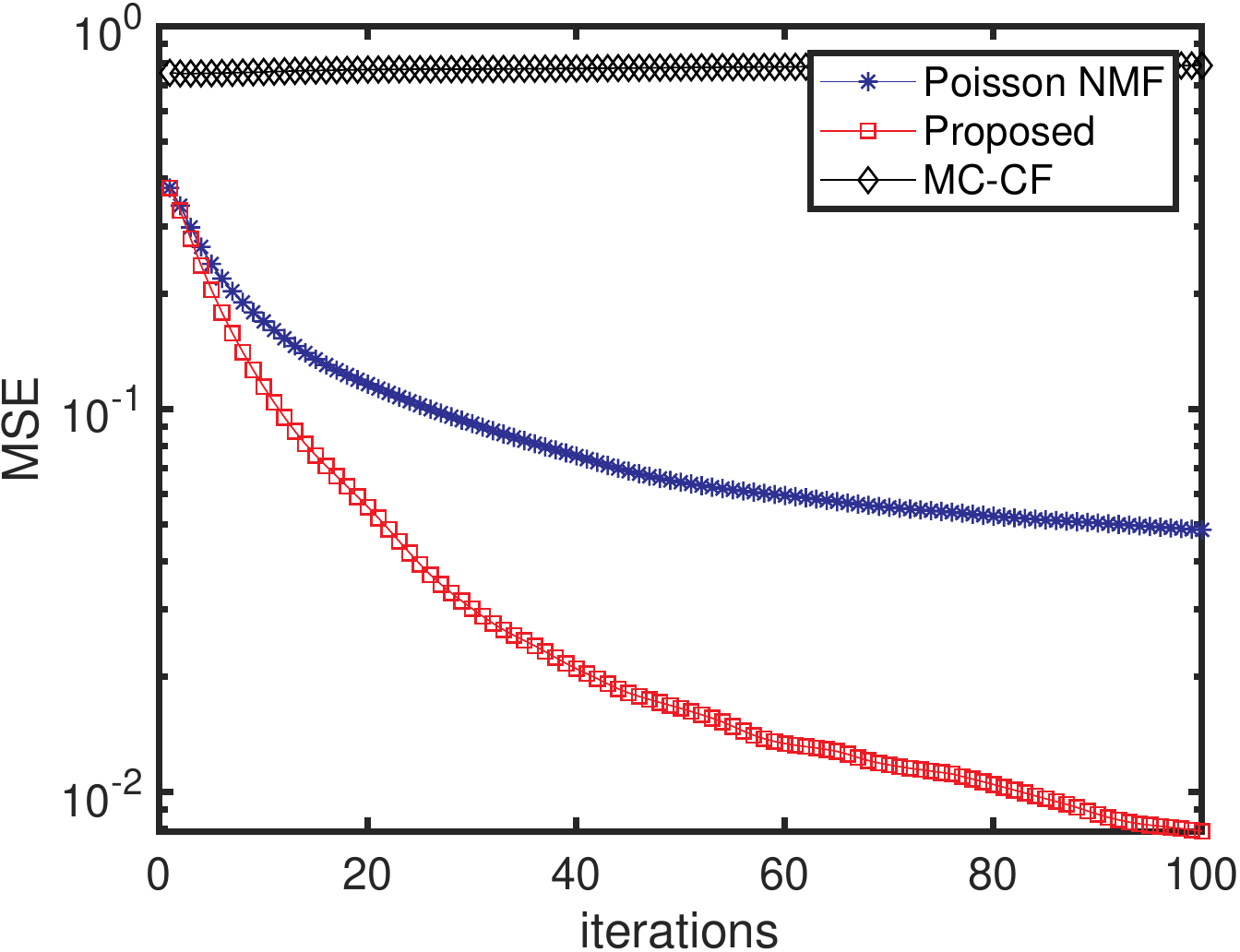}
		%	\caption{The average MSE of the estimated $\bm \alpha$ produced by the proposed algorithm.}
		%\label{fig:pollResConstAlpha}
%	\end{minipage}%
%	\begin{minipage}{.5\linewidth}
		\centering
		\includegraphics[width=.7\linewidth]{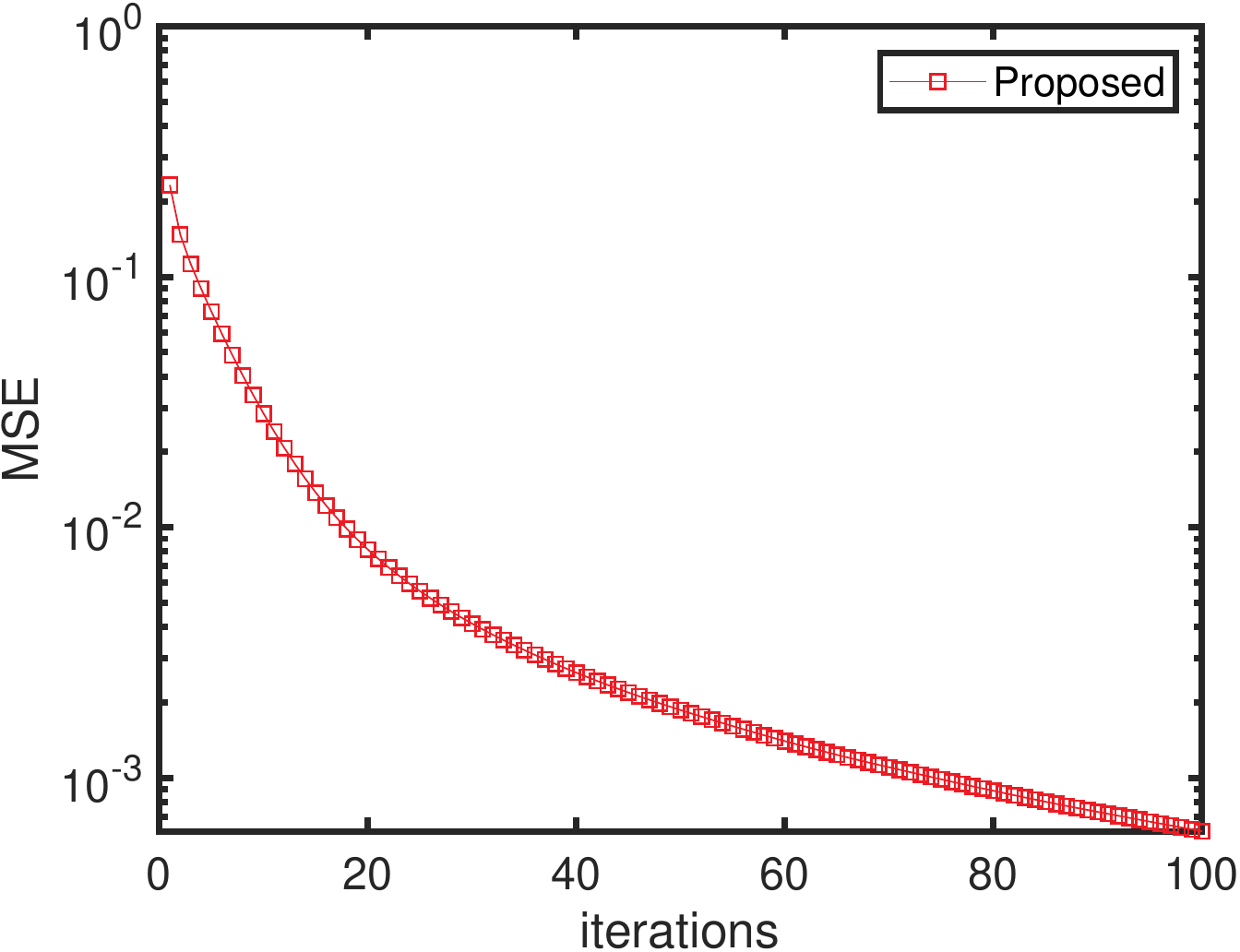}
		%	\caption{The average MSEs of the estimated $\bm \alpha$ produced by the algorithms. }
		%\label{fig:pollResFeatureAlpha}
%	\end{minipage}
	\caption{The average MSEs of the estimated $\bm U,\bm V$ (left) and $\bm \alpha$ (right) produced by the algorithms under test. }\label{fig:syn}
	%	\vspace{-.35cm}
\end{figure}

%\begin{figure}
%	\centering
%	\includegraphics[width=0.7\linewidth]{figures/iteraccvsrank}
%	\caption{The average MSEs of the estimated $\bm U,\bm V$ produced by the algorithms under different ranks of the latent NMF model. }
%	\label{fig:iteraccvsrank}
%\end{figure}

Fig.~\ref{fig:syn} (left) shows the averaged MSEs of the estimated $\bm U$ and $\bm V$ of a case where $R=8$, $I=J=50$ and $F=15$. 
In this simulation we set $\gamma=15$.
The results are averaged over 50 random trials. One can see that MC-CF does not work for this data model, since it is not specialized for integer data and does not take into consideration the imperfect detection effect. The Poisson NMF algorithm works to a certain extent---since it is a natural choice for count data as in our model. Nevertheless, the estimation accuracy is not ideal. 
The proposed algorithm exhibits the highest estimation accuracy because it explicitly accounts for imperfect detection. 
In particular, when the number of iteration reaches 100, the MSE output by the proposed method is all most one order of magnitude lower relative to Poisson NMF. This shows that considering the imperfect detection effect explicitly is quite helpful for model identification.
This result also shows that our optimization strategy is quite effective in handling the hard ML optimization problem.

Fig.~\ref{fig:syn} (right) shows the average MSE of $\bm \alpha$ for the same experiment. One can see that the accuracy increases along with the iterations---and merely using 100 iterations achieves quite a satisfactory estimation accuracy for this case.
This is encouraging, and indicates that identifying the underlying NMF model together with the selection model is possible, despite of the complex nature of this generative model.
%More synthetic experiments can be found in Appendix~\ref{app:more}.

\begin{figure}
	\centering
	%	\begin{minipage}{.485\linewidth}
	\centering
	\includegraphics[width=.7\linewidth]{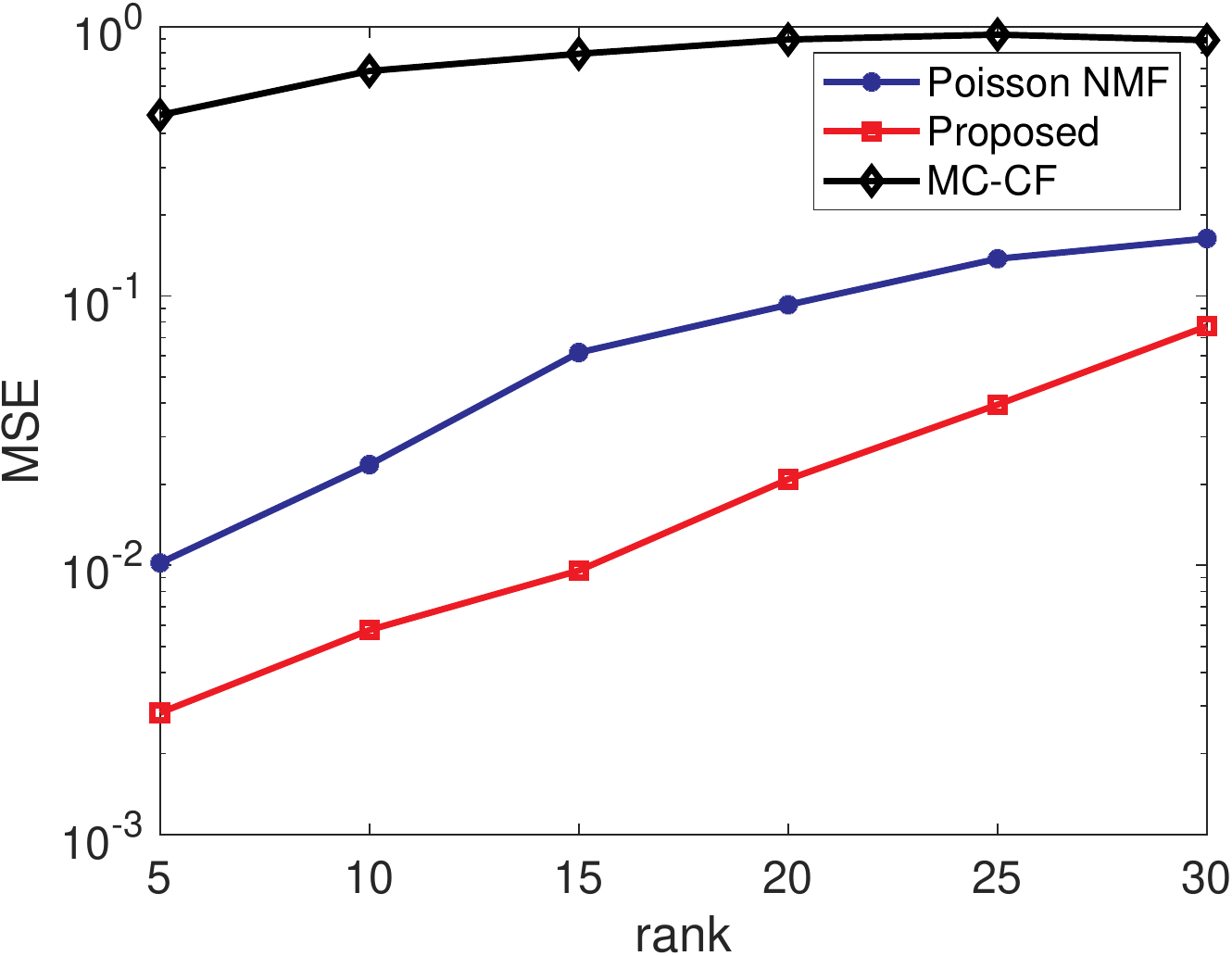}
	%	\caption{The average MSE of the estimated $\bm \alpha$ produced by the proposed algorithm.}
	%\label{fig:pollResConstAlpha}
	%	\end{minipage}%
	%	\begin{minipage}{.485\linewidth}
	\centering
	\includegraphics[width=.7\linewidth]{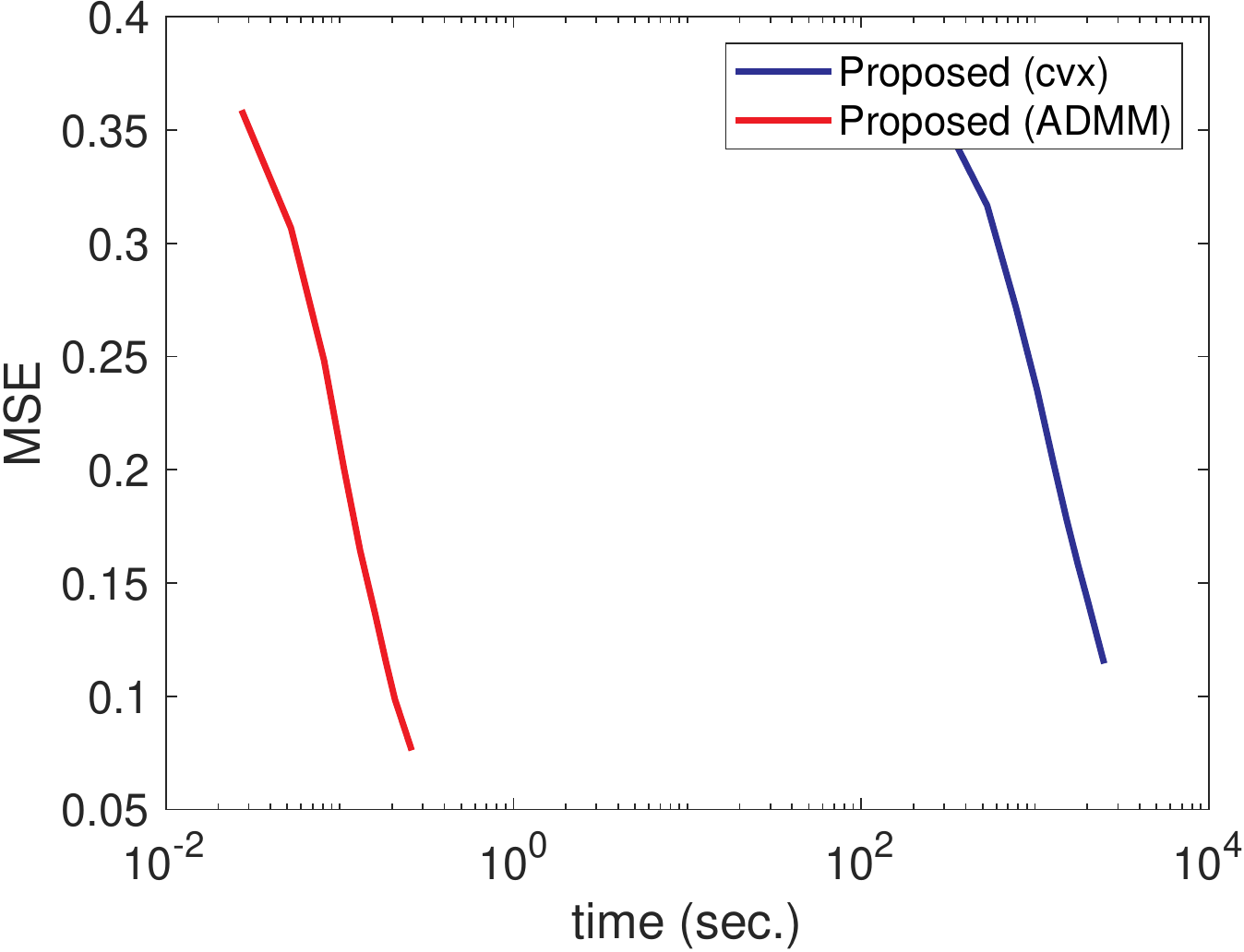}
	%	\caption{The average MSEs of the estimated $\bm \alpha$ produced by the algorithms. }
	%\label{fig:pollResFeatureAlpha}
	%	\end{minipage}
	\caption{Left: the average MSEs of the estimated $\bm \alpha$ under different ranks. Right: Runtime performance of the proposed algorithm using ADMM and CVX, respectively. }\label{fig:moreexp}
\end{figure}

Fig.~\ref{fig:moreexp} (left) shows the estimation accuracy of latent factors $\bm U,\bm V$ when the rank changes under $I=J=50$. The performance of all the algorithms under test deteriorate gracefully when $R$ increases -- which is understandable since for fixed data size, increasing the rank of $\bm U\bm V^\T$ means increasing the number of unknown parameters, and thus increasing the problem difficulty. Nevertheless, the proposed algorithm outperforms the benchmarks under all ranks under test. 

Fig.~\ref{fig:moreexp} (right) shows the efficiency of the proposed ADMM-based $\alpha$-update. The setting is the same as that use in Fig.~\ref{fig:syn}.
One can see that the ADMM algorithm outperforms a generic convex optimization solver by a very large margin, because the updates in the ADMM algorithm are all simple operations.

%Fig.~\ref{fig:iteraccvsrank} shows the estimation accuracy of latent factors $\bm U,\bm V$ when the rank changes under $I=J=50$. The performance of all the algorithms under test deteriorate gracefully when $R$ increases -- which is understandable since for fixed data size, increasing the rank of $\bm U\bm V^\T$ means increasing the number of unknown parameters, and thus increasing the problem difficulty. Nevertheless, the proposed algorithm outperforms the benchmarks under all ranks under test. 

%\subsection{Pollination Data}
\subsection{Ecological Network Data}

\begin{figure}
	\centering
	\includegraphics[width=0.76\linewidth]{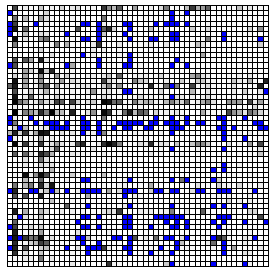}
	\caption{Visualization of the real pollination data subset we analyzed. Plants are rows, pollinators are columns, and blue entries means interaction between the corresponding pairs never co-occurred in space/time. Gray level of the grids are used to represent the number of observed counts.}
	%	\vspace{-.25cm}
	\label{fig:datasubsetvis}
\end{figure}

\noindent
{\bf Plant-Pollinator Network Dataset.} We focus here on data collected by undergraduate researchers at the H.J. Andrews (HJA) Long-term Ecological Research site in the Oregon Cascade mountains, which are publicly available \cite{Jones2017}. These data are comprised of observations at about 12 meadows over seven years. Students record every visit they observe from a pollinator to a flower---688 pollinator species and 148 plant species so far. The data are sparse; only about 3.5\% of all possible links have ever been observed. The visitation counts are also highly skewed, with many links observed only once and a few observed over 1,000 times. The recorded zeros in the data are a mix of true zeros and links that likely occurred but were unobserved during the students' watches. 
%These data form a sparse, zero-inflated tensor of highly-skewed visit counts (see Fig.~\ref{fig:datasubsetvis}). 

These network data are accompanied by {\it side information} of varying types. First, before recording the visits, the students exhaustively survey the observation plots and record the number of stalks per flowering plant species present.
%; this captures differences in context based on floral abundance across different observation periods. The students also record aspects of the observation conditions, like time of day, cloud cover, and wind, and they have assembled data on meadow characteristics like slope, size, and soil moisture. Finally, 
In addition, there are data describing traits of the species themselves, like flower color and pollinator tongue length. 

In this analysis, we focused on the most common species observed in the visitation data for which we also had information about species traits. Based on commonness and trait information, we constructed a 50 $\times$ 50 plant-pollinator visitation matrix for analysis by aggregating over all observations from 2011 through 2017 (Fig.~\ref{fig:datasubsetvis}). For these common species, 42\% of the entries were non-zero. Among the zero entries, we used the flower survey data to distinguish between two cases of zeros: 1) the plant and pollinator species occurred in the same time and place at least once but did not interact (49\% of the matrix; white cells in Fig.~\ref{fig:datasubsetvis}), vs. 2) the plant and pollinator species never co-occurred (9\% of the matrix; blue cells in Fig.~\ref{fig:datasubsetvis}). We treated the second case as truly missing values. For the first kind of zeros, we leave it to our probabilistic model to determine which are true zeros and which may have been undercounted. 
%Note that although our model does not need to know if a zero entry is truly missing or simply undercounted, we use such information here for the purpose of evaluation.
We used six categorical features in $\bm z^{(ij)}$ based on species traits: pollinator energy requirements (7 values), pollinator tongue length (8 values), plant soil community (8 values), plant life form (2 values), plant flower form (2 values), and plant exclusion platform type (whether it excluded pollinators based on its shape; 3 values). 
%Also note that incorporating held-out entries requires minor modifications to the algorithm. To be specific, we keep the same $\bm \alpha$-update as before, but only applied to the available entries. For the $\bm U,\bm V$-updates, we first impute the held-out entries $\widehat{y}_{i,j}=\widehat{\bm u}_i^\T\widehat{\bm v}_j$ (where $\widehat{\bm u}_i$, $\widehat{\bm v}_j$ denote the current estimations for $\bm u_i$ and $\bm v_j$, respectively) and then apply the proposed algorithm. This simple heuristic admits an expectation-maximization interpretation in the NMF literature \cite{kim2009weighted}.

%{\blue what are the feature vectors in our experiments? (the $\bm z^{(ij)}$'s)}

\smallskip

\noindent
{\bf Host-Parasite Dataset} We also use another ecological dataset, v.i.z., the host-parasite networks that is a part of the Sevilleta Long-Term Ecological Research program. The networks are also publicly available \cite{Dallas2017}. These data are collected over 5 years at six sites, resulting in 22 host species and 87 parasite species except for host species with less than five observations during the sampling period. These data also form a sparse and highly skewed visit count graph; about 13.5\% of all possible links have been observed. 

These host-parasite network data also contain host and parasite traits as side information. Host traits include life history traits (e.g., host diet breadth, body mass, home range size, maximum age, and species abundance) and some phylogenetic information. Parasite traits consist of life history, transmission modes, genus, type, and location.

Similar to the plant-pollinator dataset, we focus on the most common 19 host and 49 parasite species observed for analysis. For these species, 22\% of the entries are non-zero. Unlike the pollination data, the host-parasite dataset does not have additional information to distinguish ``zeros'' or ``truly missing entries''. %Hence, no held-out entries are considered in this case.
%Unlike the pollination data, the host-parasite dataset does not have additional information to distinguish ``zeros'' or ``truly missing entries''. For this reason, we assume that all zeros in the host-parasite dataset are true zeros.% for evaluation purpose. 

\smallskip

\noindent
{\bf Semi-Real Data Evaluation.} While the primary scientific objective for pollination networks is inference of the latent counts $\hat{\bm N}$ from the observations $\bm Y$, we do not have ground truth for $\bm N$ against which to compare our predictions. Instead, we employ two other evaluation strategies that are possible with the data at hand. First, we ``observe'' a real pollination data matrix according to our detection model with known parameters. This essentially replicates the simulated data evaluation above with real count networks acting as $\bm N$. The point here is that real counts $\bm N$ are not likely to strictly follow the Poisson distribution, so this analysis asks whether or not our Poisson modeling is useful in practice. The rationale is as follows:  If our two-layer model holds, the observed counts are Poisson distributed with parameter $p_{ij}\lambda_{ij}$ (also see Lemma~\ref{lem:p}). Our experiments here essentially treat the real observed counts as $\bm N$ in our model and impose another layer of Binomial detection. If we can accurately identify the imposed the detection model (i.e., accurately estimating $\bm \alpha$), it implies that the Poisson modeling for the counts is quite promising.

%Note that by Lemma~\ref{lem:p}), the observed counts, even under Bernoulli detection, should still follow a Poisson distribution, if our model holds.

\begin{figure}
	\centering
%	\begin{minipage}{.485\linewidth}
		\centering
		\includegraphics[width=.75\linewidth]{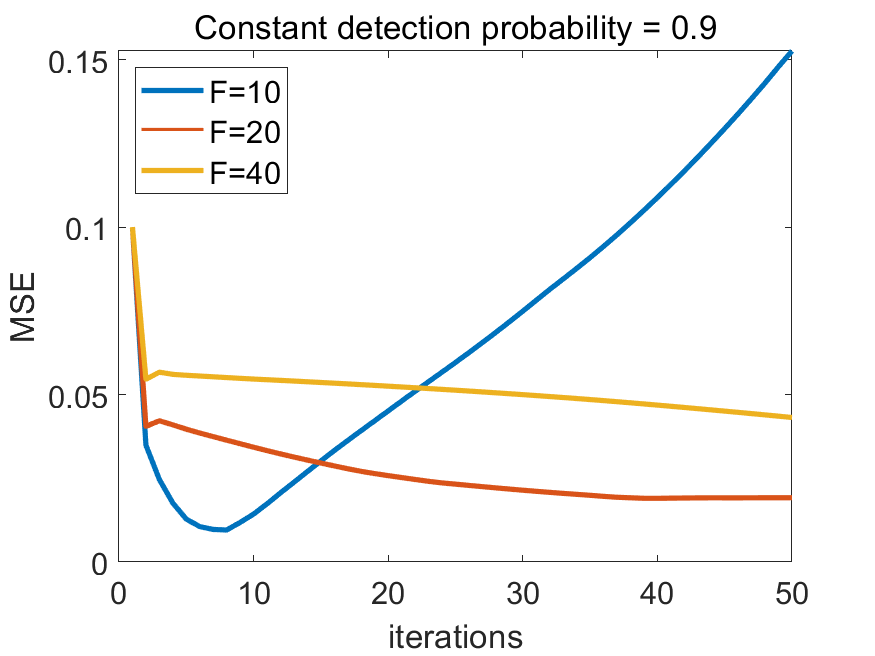}
		%	\caption{The average MSE of the estimated $\bm \alpha$ produced by the proposed algorithm.}
		%\label{fig:pollResConstAlpha}
%	\end{minipage}%
%	\begin{minipage}{.485\linewidth}
		\centering
		\includegraphics[width=.75\linewidth]{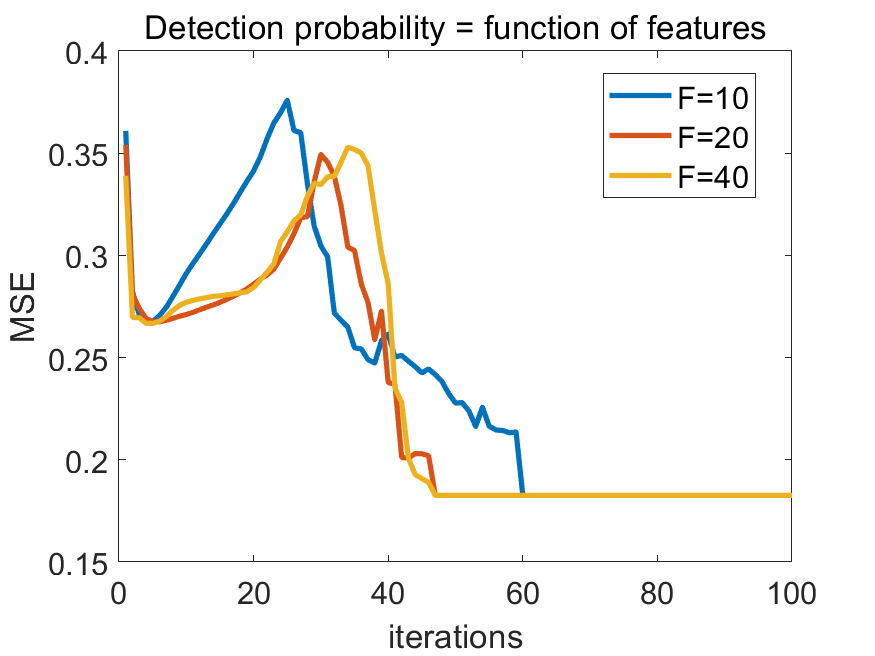}
		%	\caption{The average MSEs of the estimated $\bm \alpha$ produced by the algorithms. }
		%\label{fig:pollResFeatureAlpha}
%	\end{minipage}
	\caption{The average MSEs of the estimated $\bm \alpha$ produced by the algorithms in the plant-pollinator network. Left: constant detection probability. Right: feature-dependent detection probability. }\label{fig:semi}
	%\vspace{-.5cm}
\end{figure}

\begin{figure}
	\centering
	%	\begin{minipage}{.485\linewidth}
	\centering
	\includegraphics[width=.75\linewidth]{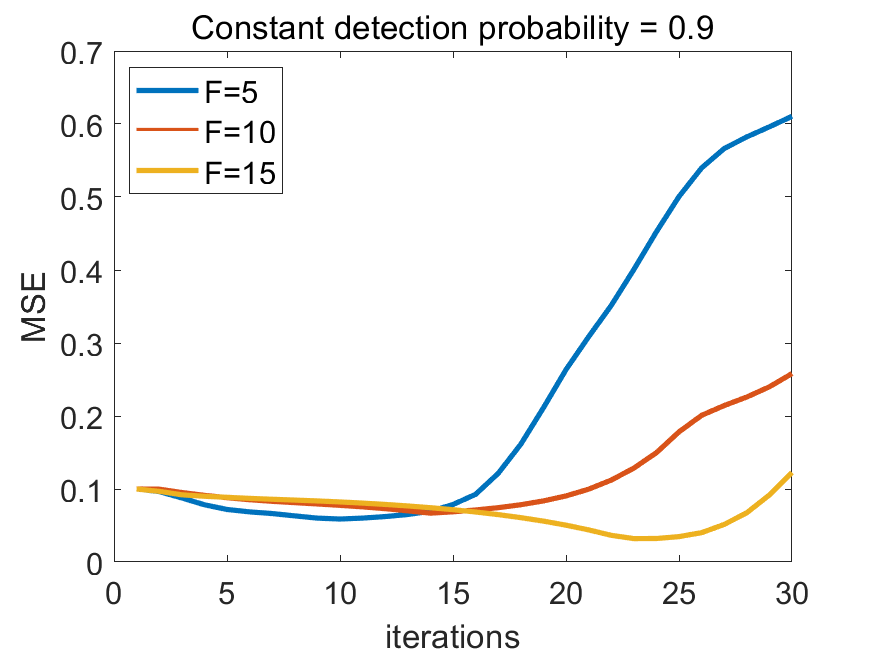}
	%	\caption{The average MSE of the estimated $\bm \alpha$ produced by the proposed algorithm.}
	%\label{fig:pollResConstAlpha}
	%	\end{minipage}%
	%	\begin{minipage}{.485\linewidth}
	\centering
	\includegraphics[width=.75\linewidth]{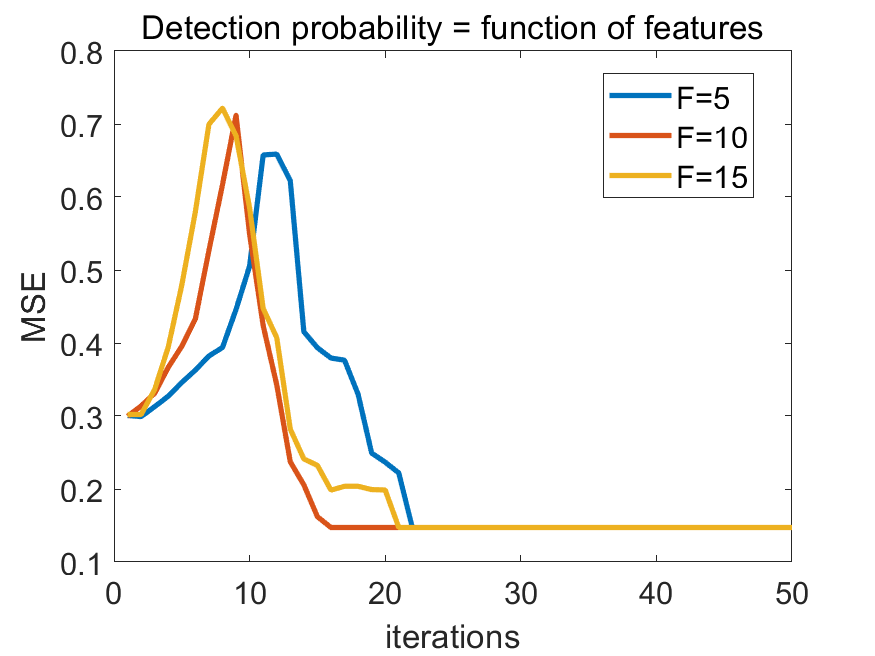}
	%	\caption{The average MSEs of the estimated $\bm \alpha$ produced by the algorithms. }
	%\label{fig:pollResFeatureAlpha}
	%	\end{minipage}
	\caption{The average MSEs of the estimated $\bm \alpha$ produced by the algorithms in the host-parasite network. Left: constant detection probability. Right: feature-dependent detection probability. }\label{fig:semi_hpi}
	%\vspace{-.5cm}
\end{figure}

We test the algorithms under two settings. First, we use a constant detection probability for all entries. Second, we generate $p_{ij}=\bm \alpha^\T {\bm z}^{(i,j)}$ using real traits ${\bm z}^{(i,j)}$ and a random vector $\bm \alpha$; the generating process is carefully controlled such that $p_{ij}\in[0,1]$.
When the detection model consists of a constant detection probability of 0.9, the model learns this parameter with generally low MSEs [see Fig.~\ref{fig:semi}, \ref{fig:semi_hpi} (upper)], if the rank is appropriately selected.% (through cross-validation). %\reminder{how did we do validation?? \blue From a validation set we find the optimal rank, and use the rank to evaluate prediction accuracy on a test set.}
For the second case, one can see that the estimation of $\bm \alpha$ is even more robust to the change of ranks [see Fig.~\ref{fig:semi}, \ref{fig:semi_hpi} (lower)], perhaps because the second case is more consistent with the assumed generative model. 
Note that the MSE of the estimated $\bm \alpha$ is not monotonic along the iterations, which is understandable since BCD algorithms only ensure monotonic decrease of the objective function.
These results indicate that the Poisson modeling in the proposed framework is plausible.
In addition, the results here also suggest that selecting an appropriate rank $F$ and using a good stopping criterion are important.

%\begin{figure}
%	\centering
%	\includegraphics[width=0.7\linewidth]{figures/const_alpha_rmsePNG}
%	\caption{The average MSE of the estimated $\bm \alpha$ produced by the proposed algorithm.}
%	\label{fig:pollResConstAlpha}
%\end{figure}
%
%\begin{figure}
%	\centering
%	\includegraphics[width=0.7\linewidth]{figures/alpha_rmsePNG}
%	\caption{The average MSEs of the estimated $\bm \alpha$ produced by the algorithms under different ranks of the latent NMF model. }
%	\label{fig:pollResFeatureAlpha}
%\end{figure}

\noindent
{\bf Real Data Evaluation.} 
The second evaluation strategy is a more typical approach in collaborative filtering: we split the matrix elements into ten folds and compare the true $\bm Y$ with $\hat{\bm Y}$ on held-out elements. We compute relative root mean squared error (${\rm rRMSE}={\rm RMSE}/\overline{Y}$, where $\overline{Y}$ is the mean value of $\bm Y$), area under the receiver operating characteristic curve (AUROC), and area under the precision-recall curve (AUPRC) on predictions for the non-missing elements aggregated across the ten test folds. 
Since the pollination network has some truely missing entries, we use the EM variation mentioned in Remark~\ref{rmk:imputation} for this dataset.
We compare against four other matrix factorization methods: Poisson NMF, implicit feedback matrix factorization (IFMF) \cite{Hu2008}, MC-CF, truncated singular value decomposition (SVD). We used performance on a validation fold as a stopping criterion for the iterative methods and ran each method with five different ranks ($F \in \{2,5,10,20,40\}$). The proposed method uses the species traits as detection features in the binomial model. 
We also use a competitive baseline, namely, the recently developed \textit{neural network-based collaborative filtering} (NCF) \cite{He2017NCF} as a benchmark.
We also compare against Poisson regression \cite{marschner2018glm}, boosted regression trees (BRT; with Poisson link) \cite{greenwell2019gbm} and random forests (RF) \cite{liaw2018rf} which predict the same ten folds without using the matrix structure. For NCF model, we tested the batch size of [256, 512], the layer-wise neuron setup of [8, 16, 32, 64] and [16, 32, 64, 128] and the learning rate of [0.0001, 0.0005, 0.001, 0.005] to tune hyper-parameters. The tree-based methods were tuned among $\{1000,2500,5000,10000\}$ trees. These baselines speak to the utility of the trait data.

\begin{table}	
	\caption{Performance of eight methods predicting held-out interactions in the plant-pollinator network data. Rank and no.of trees were tuned on a validation fold to minimize rRMSE; reporting majority vote of rank and no. trees and average iterations over 10 folds. Methods above the double line employ some form of matrix factorization; methods below the double line predict interactions solely based on trait features.}
	\centering
	\begin{tabular}{|l|c||c|c|c|}		
		\hline
		Method & Rank & rRMSE & AUROC & AUPRC \\
		& & or Trees & & \\
		\hline
		\hline
		Pois. N-mix. & 5 & \bf{3.873} & \bf{0.750} & \bf{0.733} \\
		\hline
		Pois. NMF & 5 & 9.038 & 0.735 & 0.709 \\
		\hline		
		MC-CF & 2 & 5.939 & 0.622 & 0.644 \\
		\hline
		IFMF & 2 & 7.444 & 0.631 & 0.626 \\
		\hline
		Trunc. SVD & 2 & 15.349 & 0.526 & 0.535 \\
		\hline
		\hline
		NCF & n/a & n/a & 0.731 & 0.695 \\		
		\hline		
		Pois. Regr. & n/a & 10.241 & 0.588 & 0.524 \\
		\hline
		BRT & n/a & 10.187 & 0.603 & 0.537 \\
		\hline
		RF & n/a & 10.308 & 0.661 & 0.574\\
		\hline
	\end{tabular}
	\label{tab:10foldResults}
\end{table}

\begin{table}	
	{
%	\vspace*{0.2cm}
	\caption{Performance of eight methods predicting held-out interactions in the host-parasite network data.}}
	\centering
	\begin{tabular}{|l|c|c||c|c|c|}
		\hline
		Method & Rank & rRMSE & AUROC & AUPRC \\
		& & or Trees & & \\
		\hline
		\hline
		Pois. N-mix. & 5 & \bf{3.538} & \bf{0.759} & \bf{0.692} \\
		\hline
		Pois. NMF & 10 & 3.781 & 0.707 & 0.635 \\
		\hline		
		MC-CF & 5 & 3.644 & 0.665 & 0.627 \\
		\hline
		IFMF & 2 & 3.872 & 0.639 & 0.587 \\
		\hline
		Trunc. SVD & 2 & 5.335 & 0.462 & 0.451 \\
		\hline
		\hline
		NCF & n/a & n/a & 0.689 & 0.393 \\
		\hline
		Pois. Regr. & n/a & 5.877 & 0.676 & 0.370 \\
		\hline
		BRT & n/a & 5.905 & 0.697 & 0.383 \\
		\hline
		RF & n/a & 5.920 & 0.663 & 0.343\\
		\hline
	\end{tabular}
	\label{tab:host-parasite}
	
	%	\vspace{-.3cm}
\end{table}

The results show that the proposed approach outperforms all competitors on all three metrics (Table \ref{tab:10foldResults}). Among the baselines, Poisson NMF have the most similar performance as that of the proposed method. 
This is not entirely surprising---as we have seen in Lemma~\ref{lem:key}, if $\bm N$ is Poisson distributed, then $\bm Y$ is also Poisson distributed, given that the observation is truly a Binomial selection process.
Nevertheless, the proposed method still performs better compared to a simple Poisson NMF formulation, even for the task of predicting $\bm Y$.
The reason might be that the proposed approach can effectively incorporate trait features, which should be helpful in practice.
The proposed model also outperforms NCF which is known to be effective for collaborative filtering by replacing the inner product with a neural architecture. It shows that capturing hidden interaction mechanisms from the sampled counts in the field, which could have many missing data points unlike user feedback, can be a challenging problem to user-oriented recommendation models. In this regard,
%In addition, 
the proposed approach can predict additional information in the form of latent network counts, i.e., $\bm N$ (to be discussed shortly), which is more important in ecology. 
%That BRT and RF outperform Poisson regression indicates an advantage of allowing nonlinear relationships to the trait information. 
%{\blue That BRT outperforms Poisson regression indicates an advantage of allowing nonlinear relationships to the trait information.}
%We will consider allowing such flexibility within our generative framework in future work.

As a third evaluation strategy, we examine predictions made from our model for the true missing entries in the pollination network (blue in Fig.~\ref{fig:datasubsetvis}). We sort the 226 missing entries from largest to smallest predicted interaction counts. We showed the top ten and bottom ten interactions to an expert entomologist and asked which set were more likely to interact if they did co-occur (without revealing our predictions). The expert chose the set of the top ten interactions as more likely than the bottom ten, in part because the bottom ten included some plants with exclusionary characteristics that would forbid some insects from visiting them. While it is challenging to get quantitative validation, this qualitative feedback is encouraging and also very intriguing.  

Finally, we examine differences in the $\hat{\bm N}$ estimated from pollination data with the observed $\bm Y$. As mentioned above, only 42\% of the possible links in the pollination network were observed. In contrast, 80.1\% of the links were estimated to have a latent count greater than zero. This implies that the data maybe only recorded roughly half of the actual links in the network. This effect of imperfect detection extends to other network statistics of ecological importance as well. One potential application of these estimates  $\hat{\bm N}$ is that they can be offered to ecologists as references for designing and taking field observations.

\section{Conclusion and Discussion}
In this work, we proposed a two-layer statistical model for network analysis in ecology.
Our model captures a key aspect of many ecological networks (e.g., pollination networks)---the interaction counts between species are usually systematically undercounted, which makes existing collaborative filtering approaches inapplicable for such networks. 
We proposed a generative model that is a judicious integration of Poisson low-rank latent matrix factorization and Binomial selection,
and we devised an effective optimization algorithm to handle the associated challenging maximum likelihood model identification problem.
We evaluated the proposed method on both synthetic and real data. Excitingly, evaluation on the real pollination network shows that the proposed model and algorithm are promising.

As future work, one possible extension is to introduce a richer model for the Binomial selection stage.
Our current model uses a linear regression model, which strikes a good balance between simplicity and effectiveness. However, a nonlinear regression model using kernels or neural nets may capture the reality even better, thereby further boosting performance.
%In addition, the observation model needs not to be Binomial. Many other models that are used in Ecology ({\blue e.g., XXX}) can be considered.
In addition, we plan to develop a version of this modeling framework that allows both latent \textit{and} observed features to inform the network model. This would allow the trait data to influence the interaction counts directly rather than through the effects of imperfect detection, leaving the latent features to capture unmeasured aspects of the propensity for interactions to occur.

%\noindent
%{\bf Acknowledgment.} This work is supported in part by the National Science Foundation under projects NSF-ECCS 1608961 and NSF-ECCS 1808159. The authors would like to thank entomologist Andrew Moldenke for expert advice.

%\bibliographystyle{ACM-Reference-Format}

\bibliography{refs_xiao_meta}

\begin{IEEEbiography}[{\includegraphics[width=1in,height=1.25in,clip,keepaspectratio]{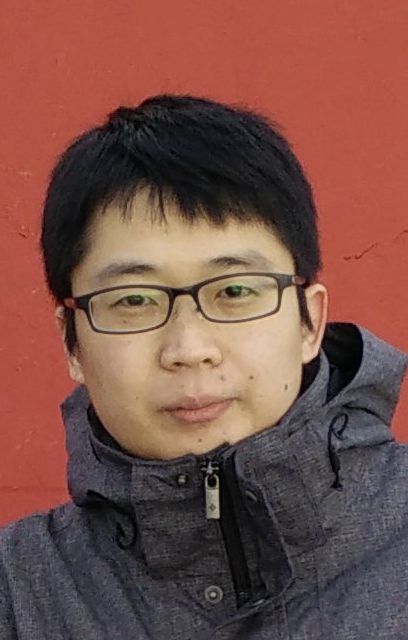}}]
	{Xiao Fu} (S'12-M'15) is an Assistant Professor in the School of Electrical Engineering and Computer Science, Oregon State University, Corvallis, Oregon, United States. He received his Ph.D. degree in Electronic Engineering from The Chinese University of Hong Kong (CUHK), Hong Kong, 2014. He was a Postdoctoral Associate in the Department of Electrical and Computer Engineering, University of Minnesota, Minneapolis, MN, United States, from 2014-2017. His research interests include the broad area of signal processing and machine learning. He received a Best Student Paper Award at ICASSP 2014, and co-authored a Best Student Paper Award at IEEE CAMSAP 2015.
\end{IEEEbiography}

\vspace{-1cm}
\begin{IEEEbiography}[{\includegraphics[width=1in,height=1.25in,clip,keepaspectratio]{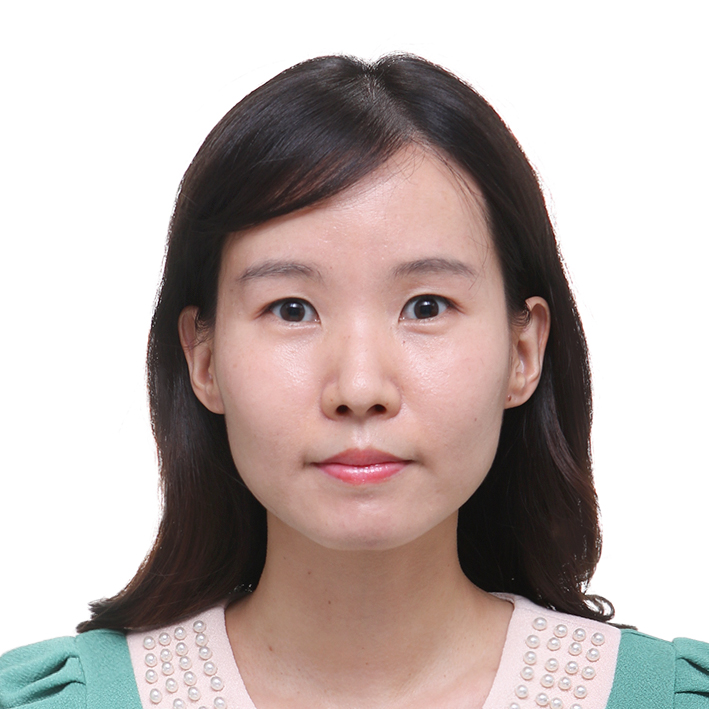}}]
	{Eugene Seo} received her B.Eng. degree in Computer Science and Electronic Engineering from Handong Global University, South Korea, in 2007, and M.S. degree in Computer Science from Korea Advanced Institute of Science and Technology, South Korea, in 2011.
	She is currently a Ph.D. candidate student at the School of Electrical Engineering and Computer Science, Oregon State University, Corvallis, Oregon, United States. 
	Her research interests include artificial intelligence, machine learning, data mining, recommender systems, and computational sustainability.
\end{IEEEbiography}

\vspace{-1cm}
\begin{IEEEbiography}[{\includegraphics[width=1in,height=1.25in,clip,keepaspectratio]{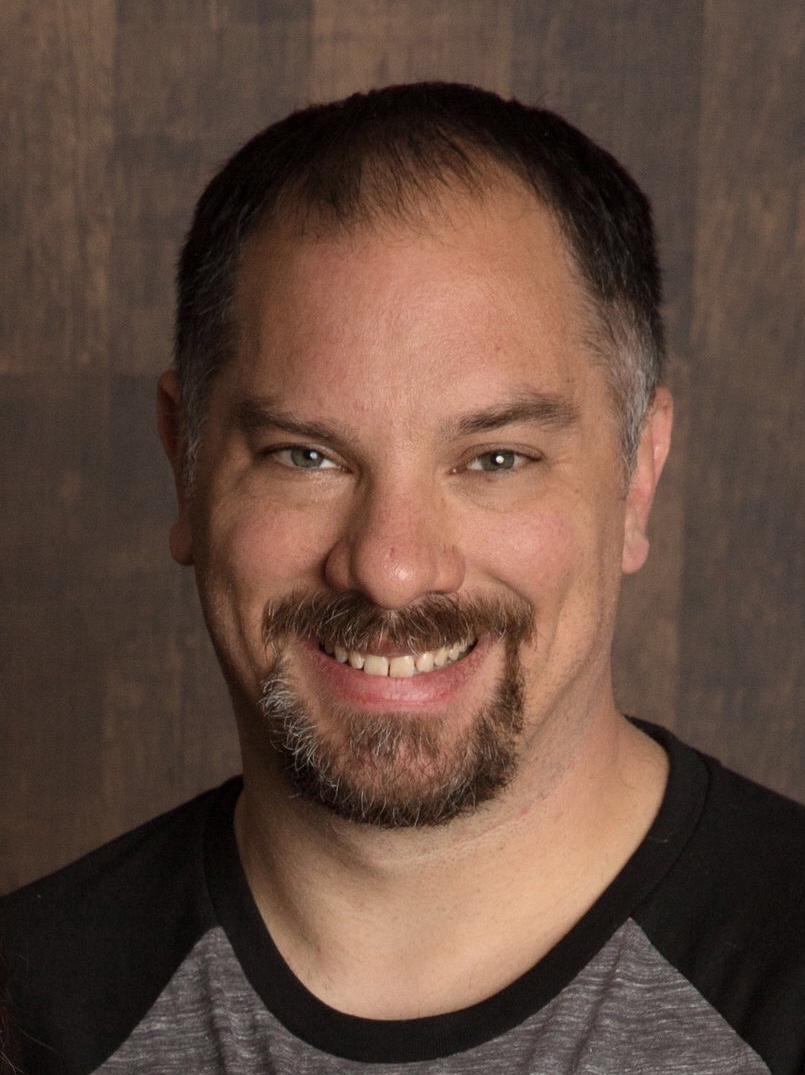}}] {Justin Clarke}
	is a senior at Oregon State University majoring in Computer Science. He is currently working as an intern at the NASA Ames Research Center in the Robust Software Engineering group. This fall he will join the Ph.D. program at the University of Massachusetts, Amherst. His research interests include machine learning and causal modeling. 
\end{IEEEbiography}

\vspace{-1cm}
\begin{IEEEbiography}[{\includegraphics[width=1in,height=1.25in,clip,keepaspectratio]{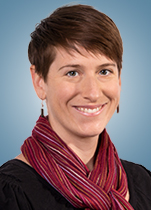}}]
	{Rebecca A. Hutchinson} received her B.S. degree in Computer Science and Engineering from Bucknell University, Lewisburg, PA, United States, in 2002.
	She received her Ph.D. degree in Computer Science from Carnegie Mellon University, Pittsburgh, PA, United States, in 2009.
	From 2009 to 2015 she has been a Post Doctoral Scholar and Fellow with the School of Electrical Engineering and Computer Science, Oregon State University, Corvallis, Oregon, United States.
	She is currently a Assistant Professor in the School of Electrical Engineering and Computer Science and the Department of Fisheries and Wildlife, Oregon State University, Corvallis, Oregon, United States.    
	Her research is at the intersection of machine learning and ecology, including computational sustainability, species distribution modeling, hierarchical latent variable models, and robust parameter estimation methods.
\end{IEEEbiography}

\appendices

%{\Large Supplementary Materials for ``Link Prediction Under Imperfect Detection: Collaborative Filtering for Ecological Networks''}

\section{Proof of Lemma~\ref{lem:key}}
\label{app:infSumDeriv}
The derivation is a classic result in statistics \cite{dennis2015computational}.
To obtain the simple expression in \eqref{eq:simplified}, let us begin with the following:
\begin{align}
&\sum_{n=y_{ij}}^\infty \frac{\lambda_{ij}^{n} e^{-\lambda_{ij}}}{n!}  \frac{n!}{y_{ij}!(n-y_{ij})!} p^{y_{ij}} (1-p)^{n-y_{ij}}\\
& = \frac{p^{y_{ij}}e^{-\lambda_{ij}}}{y_{ij}!} \sum_{n=y_{ij}}^\infty \frac{\lambda_{ij}^{n}(1-p)^{n-y_{ij}}}{(n-y_{ij})!}.
\end{align}
Now, consider the following change of variables:
$ m_{ij} = n-y_{ij}.$
Then, what we have is
\begin{align*}
&\sum_{n=y_{ij}}^\infty \frac{\lambda_{ij}^{n} e^{-\lambda_{ij}}}{n!}  \frac{n!}{y_{ij}!(n-y_{ij})!} p^{y_{ij}} (1-p)^{n-y_{ij}}\\
& = \frac{p^{y_{ij}}e^{-\lambda_{ij}}}{y_{ij}!} \sum_{n=y_{ij}}^\infty \frac{\lambda_{ij}^{n}(1-p)^{n-y_{ij}}}{(n-y_{ij})!}\\
& = \frac{p^{y_{ij}}e^{-\lambda_{ij}}}{y_{ij}!} \sum_{m_{ij=0}}^\infty \frac{\lambda_{ij}^{m_{ij}+y_{ij}}(1-p)^{m_{ij}}}{m_{ij}!}\\
& = \frac{(p\lambda_{ij})^{y_{ij}}e^{-\lambda_{ij}}}{y_{ij}!} \sum_{m_{ij=0}}^\infty \frac{\lambda_{ij}^{m_{ij}}(1-p)^{m_{ij}}}{m_{ij}!} \\
& =  \frac{(p\lambda_{ij})^{y_{ij}}e^{-\lambda_{ij}p}}{y_{ij}!} %\sum_{m_{ij=0}}^\infty \frac{\lambda_{ij}^{m_{ij}}(1-p)^{m_{ij}}}{m_{ij}!}
\end{align*}
where the last equality is by the Taylor expansion.

\section{Proof of Proposition~\ref{prop:convergence}}\label{app:converge}
In this section, we present the convergence proof of the proposed algorithm.
%\subsection{The BSUM Framework}
Let us first introduce the \textit{block successive upper bound minimization} (BSUM) \cite{razaviyayn2013unified,hong2016unified}, which will be applied to our case.

Consider an optimization problem
\begin{subequations}\label{eq:BCD}
	\begin{align}
	\minimize_{\x_1,\ldots,\x_n}&~f(\x_1,\ldots,\x_n)\\
	{\rm suject~to}&~\x_1\in X_1,\ldots,x_n\in X_n,
	\end{align}
\end{subequations}
where $\x_i$ denotes the $i$th block of the optimization variables and $X_i$ denotes a convex closed set.
The BSUM framework advocates the following algorithm to update $\x_i$ for $i=1,\ldots,n$ cyclically:
\begin{align}
\x_i^{t+1} &\leftarrow \arg\min_{\x_i\in X_i}~g_i(\x_i; \x_{-i}^t) \label{eq:ub}
%	\x_j^{t+1}& \leftarrow \x_j^t,~\forall j\neq i\\
%	t &\leftarrow t + 1.
\end{align}
where
\[   \x_{-i}^t =\left(\x_1^{t+1},\ldots,\x_{i-1}^{t+1},\x_{i+1}^{t},\ldots,\x_n^t \right).   \]
The BSUM framework bears a lot of resemblances to the Gauss-Seidel type \textit{block coordinate descent} (BCD) scheme \cite{bertsekas1999nonlinear}. 
The key difference lies in
the employment of \eqref{eq:ub}: Assuming both $f$ and $g_i$ are continuously differentiable,  the $g_i$ function is an \textit{optimization surrogate} that satisfies
\begin{align}\label{eq:g_i}
g_i(\x_i; \x_{-i}^t) & \geq f(\x_i; \x_{-i}^t),~\forall \x_i\in X_i\\
g_i(\x_i^t; \x_{-i}^t) & = f(\x_i^t; \x_{-i}^t),\\
\nabla_{\x_i } g_i(\x_i^t ; \x_{-i}^t) & = \nabla_{\x_i } f(\x_i^t ; \x_{-i}^t).
\end{align}
Eq.~\eqref{eq:g_i} means that $g_i$ is a blockwise tight upper bound of $f$.
It was shown in \cite{razaviyayn2013unified} that the following holds:
\begin{theorem}
	Assume that $f(\cdot)$ is differentiable with respect to all $\x_i$ for $i=1,\ldots,n$, and that the block optimization surrogate $g_i(\cdot)$ for all $i$ satisfies \eqref{eq:g_i}. Then, every limit point of the solution sequence $\{x^{t}\}$ produced by the BSUM algorithm is a stationary point of Problem~\eqref{eq:BCD}.
\end{theorem}

%\subsection{Proposed Algorithm as BSUM}
The proposed algorithm admits three blocks, i.e., $\bm U$, $\V$, and $\bm \alpha$.
let us denote $\bm x_1={\rm vec}(\bm U)$, $\x_2={\rm vec}(\bm V)$ and $\x_3=\bm \alpha$.
It is clear that $g_3(\cdot)$ satisfies \eqref{eq:g_i}, since
\[    g_3(\x_3;\x_{-3}^t) = f(\x_3;\x_{-3}^t)    \]
in our case.
In addition, for $\x_1$, we notice that Eq.~\eqref{eq:g_i} also holds by the construction in Eq.~\eqref{eq:upp}.
A subtle point is the upper bound construction holds in Eq.~\eqref{eq:upp} only when there is no zero element in $\bm U$ and $\bm V$ --- which is why Proposition~\ref{prop:convergence} has this condition assumed.
In practice, one normally has no control of intermediate iterates $\bm U^t$ and $\bm V^t$, and thus assuming this is considered relatively strong.
One pragmatic remedy is to modify the update in \eqref{eq:UV} to
\begin{align*}
&\bm U \leftarrow  \left(\bm U \circ \bm \Phi\right) / \left(\tilde{\bm U} + \epsilon \bm 1\bm 1^\T\right),~\bm \Phi =\left(\bm Y/\bm U\bm V^\T \right)\bm V\\
&\V \leftarrow \left(\bm V \circ \bm \Psi \right) / \left(\tilde{\bm V}+ \epsilon \bm 1\bm 1^\T\right),~\bm \Psi =\left(\bm Y^\T/\bm V\bm U^\T \right)\bm U,
\end{align*}
where $\bm 1\bm 1^\T$ is an all-one matrix with proper size, and $\epsilon>0$ is small number.
With this modification, $\bm U^t$ and $\bm V^t$ are always positive.
This trick has been found effective for stabilizing such multiplicative updates.

%\section{Additional Experiments}

%\section{More Experiments}\label{app:more}

\end{document}